\documentclass[12pt,a4paper,superscriptaddress,floatfix]{article}

\advance\voffset by -2.0cm \advance\hoffset by -1.25cm
\usepackage{amsmath}
\usepackage{amssymb}
\usepackage{amsthm}
\usepackage{amsfonts}

\numberwithin{equation}{section}

\theoremstyle{definition}

\newcommand{\CC}{\mathbb{C}} 
\newcommand{\RR}{\mathbb{R}} 
\newcommand{\ZZ}{\mathbb{Z}} 
\newcommand{\NN}{\mathbb{N}} 

\newcommand{\DD}{\mathbb{D}}
\newcommand{\HH}{\mathbb{H}} 

\newcommand{\be}{\begin{equation}}
\newcommand{\ee}{\end{equation}}

\def\1{\frak 1}
\def\2{\frak 2}
\def\3{\frak 3}

\newlength{\oldcolsep}

\begin{document}

\title{Wilson Loops on Riemann Surfaces, Liouville Theory\\ and Covariantization of the Conformal Group}
\author{Marco Matone and Paolo Pasti}\date{}

\maketitle

\begin{center} Dipartimento di Fisica e Astronomia ``G. Galilei'' \\
 Istituto
Nazionale di Fisica Nucleare \\
Universit\`a di Padova, Via Marzolo, 8-35131 Padova,
Italy\end{center}

\bigskip

\begin{abstract}
\noindent The covariantization procedure is usually referred to the translation operator, that is the derivative.
Here we introduce a general method to covariantize arbitrary differential operators, such as the ones defining the fundamental group of a given
manifold. We focus on the differential operators representing the ${\rm sl}_2(\RR)$ generators,
which in turn, generate, by exponentiation, the two-dimensional conformal transformations.
A key point of our construction is the recent result on the closed forms of the Baker-Campbell-Hausdorff formula. In particular, our covariantization
receipt is quite general. This has a deep consequence since it means that the covariantization of the conformal group is {\it always definite}.
Our covariantization receipt is quite general and apply in general situations, including AdS/CFT. Here we focus on  the projective unitary representations of the fundamental group
of a Riemann surface,
which may include elliptic points and punctures,
introduced
in the framework of noncommutative Riemann surfaces.
It turns out that the covariantized conformal operators are built in terms of Wilson loops around Poincar\'e geodesics, implying a deep relationship between
gauge theories on Riemann surfaces and Liouville theory.

\end{abstract}

\newpage

\section{Introduction}\label{intro}

It is well-known that in 2$D$ space-time a pure gauge theory is locally trivial. In particular, by $\partial_\mu F^{\mu\nu}=0$, it follows that the unique nontrivial component $E_1=F_{01}$ is a constant and the finite energy condition fixes $E_1=0$. This is a consequence of the fact that the number of degrees of freedom of a pure gauge theory in $D$-dimensions is $D-2$. Essentially, in topologically trivial 2D space-time any gauge configuration can be absorbed by a gauge transformation.

\vspace{.4cm}

\noindent
The situation is completely different in the case one considers non-trivial topologies. In particular, there are well-known models corresponding to a topological field theory with a strict relationship with string theory. Consider the case of a $SU(N)$ (or $U(N)$) Yang-Mills theory on a genus $g$ Riemann surface $\Sigma$
$$
{\cal Z}_\Sigma=\int DA^\mu exp\Big[-{1\over 4g^2}\int_\Sigma d^2 x\sqrt {\det g}\, {\rm tr} F_{    \mu\nu} F^{\mu \nu} \Big]  \ ,
$$
where the trace is on the fundamental representation. Cutting $\Sigma$ along a basis of its fundamental group $\pi_1(\Sigma)$, one gets the relation
$$
a_1b_1 a_1^{-1}b_1^{-1} \ldots a_gb_g a_g^{-1}b_g^{-1}=I  \ .
$$
It turns out that the partition function can be expressed in terms of path integral of the traces of matrices of the gauge group associated to each generator of $\pi_1(\Sigma)$ and (see \cite{Frishman:2010zz} and references therein)
$$
{\cal Z}_\Sigma=\sum_R d_R e^{-{g A c_2(R)\over2}}\int \prod D U_l DV_l {\rm tr}_R[U_1V_1U_1^+V_1^+\ldots
U_gV_gU_g^+V_g^+]  \ ,
$$
where the summation is over the irreducible representations of the group, $A$ is the area of $\Sigma$, $d_R$ the dimension of the representation $R$ and $c_2(R)$ the second Casimir operator of $R$.

\vspace{.4cm}

\noindent
The relation between connections and the fundamental group of $\Sigma$ appears also in the context of noncommutative Riemann surfaces. This corresponds to consider
 unitary projective representations of the uniformizing
Fuchsian groups.
Such representations are obtained by covariantizing the differential operators of the ${\rm sl}_2(\RR)$ algebra.
Roughly speaking, whereas the projective unitary representation of the (abelian) group uniformizing the torus
\begin{equation}
{\cal U}_1{\cal U}_2=e^{2\pi i\theta}{\cal U}_2{\cal U}_1 \ ,
\label{noncommtoruspre}\end{equation}
is simply obtained by setting ${\cal U}_k=\exp({\lambda_k( \partial_k+iA_k)})$, with $A$ a connection 1-form, in the case of
uniformizing groups one needs to covariantize the generators of ${\rm sl}_2(\RR)$. In other words, first one has to find the receipe
to covariantize $\partial_z$, $z\partial_z$ and $z^2\partial_z$ and then finding the unitary operators ${\cal U}_{k}$ such that
\begin{equation}
\prod_{k=1}^g{\cal U}_{2k-1}\,{\cal U}_{2k}\,{\cal
U}_{2k-1}^{-1}\,{\cal U}_{2k}^{-1}=e^{2\pi i\theta}I \ .
\label{inonniAAA}\end{equation}
This case corresponds to the one of hyperbolic Riemann surfaces. However, we will consider the general case that includes Riemann surfaces with elliptic point and punctures.

\vspace{0.4cm}

\noindent
We will see that finding the unitary projective representation connects several questions, such as the one of the simultaneous covariantization, originally considered in our previous work
\cite{Bertoldi:2000ua}\cite{Bertoldi:2000hj}, that now are investigated in a systematic way and solved step by step. In our investigation,
we will express the M\"obius transformations in terms of the differential representation of ${\rm sl}_2(\RR)$. This is done by first expressing and element of ${\rm PSL}_2(\RR)$ as the composition of a translation, dilatation and a
special conformal transformation.
However, this has a drawback for our purposes.
The reason is that the order of the above composition may change according to the kind of M\"obius transformation one is considering. This is related to the important question of expressing
${\cal U}_{k}$ as the exponential of a unique operator, conjugated
by a functional $F_k$ of the connection $A$, associated to the geodesic defining the corresponding generators of the uniformizing group, namely
\be
{\cal U}_{k}=F_k(z,\bar z)\exp(D_k)F_k^{-1}(z,\bar z) \ .
\ee
The problem then is to find $D_k$ and
the functional $F_k$, which is the key object to covariantize  $D_k$. As we will see, such questions are in turn related to the problem of finding the explicit relation between the normal form of $\pi_1(\Sigma)$ and the
uniformizing group $\Gamma$ of $\Sigma$. We will find such a relation, that selects the corresponding form of the generators $\beta_k$ of $\Gamma$. As such, these generators satisfy the relation
\be
\beta_{4g}\beta_{4g-1}\ldots \beta_1=I \ ,
\ee
which is the case of compact Riemann surfaces of genus $g$ (our investigation extends to the case with elliptic points and punctures).
\noindent As we will see, all the above questions, unanswered in \cite{Bertoldi:2000ua}\cite{Bertoldi:2000hj}, have a solution which is essentially unique.

\vspace{0.4cm}

\noindent
The operator
$\exp(D_k)$ performs the M\"obius transformation $\beta_k^{-1}$ of the arguments of a function, so that
\be
{\cal U}_{k}=F_k(z,\bar z)F_k^{-1}(\beta_k^{-1}z,\beta_k^{-1}\bar z)\exp(D_k) \ ,
\ee
where
\be
\beta_kz:={a_kz+b_k\over c_k z+d_k} \ .
\label{lammobbius}\ee
The $F_k$'s are directly related to the Wilson loop associated to the geodesic connecting $z$ and $\beta_k^{-1}z$ on the upper half plane $\HH$, namely
\be
F_k(z,\bar z)F_k^{-1}(\beta_k^{-1}z,\beta_k^{-1}\bar z)=
\exp\Big({ib\int_z^{\beta_k^{-1}z}A}\Big) \ ,
\ee
where $b$ is a real parameter. Therefore,
\be
{\cal U}_k=\exp\Big({ib\int_z^{\beta_k^{-1}z}A}\Big)\exp(D_k) \ .
\ee
Note that
\be
W_{\beta_k}=\exp\Big({ib\int_z^{\beta_k^{-1}z}A}\Big) \ ,
\ee
projects to a Wilson loop on $\Sigma$.
We call the ${\cal U}_k$'s, Wilson--Fuchs operators.
The modulus of
\be
d_A(z,w)=\int_z^w A \ ,
\ee
related to the Wilson loop by $W_{\beta_k}=\exp(ibd_A(z,\beta_k^{-1}))$, is a pseudo distance that called ``gauge-length'' in \cite{Bertoldi:2000ua}\cite{Bertoldi:2000hj}.
It corresponds to the Poincar\'e area of the hyperbolic triangle whose sides are the geodesic joining $z$ and $w$, together with the two geodesics connecting $z$ and $w$ to the point at imaginary infinity on the upper half-plane. An outcome of \cite{Bertoldi:2000ua}\cite{Bertoldi:2000hj} is that M\"obius transformations correspond to gauge transformations.
\\

\vspace{.4cm}

\noindent
As we said, the above construction is possible once one finds $D_k$ such that any M\"obius transformation can expressed in the form $\exp(D_k)$. Such a
question is equivalent to the problem of finding the closed form of $W$ such that
\be
\exp(X)\exp(Y)\exp(Z)=\exp(W) \ .
\label{solllv}\ee
The recent solution of such a problem is a key point of our construction. In particular, it turns out that our covariantization
receipt is quite general. This has a deep consequence since it means that the covariantization of the conformal group is {\it always definite}.
In \cite{Matone:2015wxa}, it has been introduced an algorithm to derive the closed form of $W$ in (\ref{solllv}) for a wide class of commutator algebras, classified in
\cite{Matone:2015xaa}
 and applied to all semisimple complex Lie algebras in \cite{Matone:2015oca}. The algorithm in \cite{Matone:2015wxa}, that extends the remarkable result by Van-Brunt and Visser
 \cite{Van-Brunt:2015ala} (see also \cite{Van-Brunt:2015bza} for related results),
exploits the associative property of the Baker-Campbell-Hausdorff formula and implementing in it the Jacobi identity.
In particular, it turns out that when $X$, $Y$ and $Z$ are elements of ${\rm sl}_2(\RR)$, the corresponding
commutator algebras is, according to the classification in \cite{Matone:2015xaa}, a subtype of the {\it type 4}.
As a result, we will see that the M\"obius transformation (\ref{lammobbius})
is represented by the unitary operator ${\cal U}_k=\exp({\cal D}_k)$, where ${\cal D}_k$ is the ${\rm sl}_2(\RR)$ covariantized operator
$$
{\cal D}_k=
{\lambda_+^{(k)}-\lambda_-^{(k)}\over e^{-\lambda_-^{(k)}}-e^{-\lambda_+^{(k)}}}\times
$$
\be
\times F_k\big[\lambda_{-1}^{(k)}(\partial_z+\partial_{\bar z})+(2-e^{-\lambda_+^{(k)}}-e^{-\lambda_-^{(k)}})
(z\partial_z+\bar z\partial_{\bar z}+1)+\lambda_1^{(k)}(z^2\partial_z+{\bar z}^2\partial_{\bar z}+z+\bar z)\big]F_k^{-1} \ ,
\label{laTgrBAAA}\ee
\\
\noindent
where the parameters $\lambda_\pm^{(k)}$ and $\lambda_j^{(k)}$, $j=-1,01$, are defined in terms of the components of the matrix $\beta_k$.
\\
\noindent
Usually, the covariantization procedure is  considered only for the translation operator, that is for the ordinary derivative. The above covariantization is the one for the
conformal transformations.
To derive (\ref{laTgrBAAA}) it has been introduced a general method
to covariantize much more general operators than $\partial_z$. Actually, our analysis starts with the torus, whose fundamental domain are straight lines, and these are generated by derivatives.
The noncommutative torus leads to a covariantization of the derivatives that apparently cannot be extended to the higher genus case. However, we reformulate the
 noncommutative torus in a more geometrical way. It turns out that such a geometrical formulation is the natural framework to derive, in analogy with the case of the torus, the corresponding quantities
 for negatively curved Riemann surfaces. Such a strategy provides the correct prescription to define the covariantized operators and leads to (\ref{laTgrBAAA}).

\noindent The above covariantization prescription can be extended to more general cases, including AdS/CFT. In particular whenever
a manifold is naturally associated to a differential representation of its fundamental group.
Here we focus on  the projective unitary representations of the fundamental group
of a Riemann surface,
which may include elliptic points and punctures,
introduced
in the framework of noncommutative Riemann surfaces.

\vspace{0.4cm}

\noindent The organization of the paper is as follows. In section 2, after shortly reviewing the uniformization theorem, we express the M\"obius transformations in terms of the exponentiation of differential operators representing ${\rm sl}_2(\RR)$. In section 3 we express the action of such operators in the form of the exponential of a linear combination of the ${\rm sl}_2(\RR)$ generators. This is a basic step to formulate the covariantization of the conformal group.
In section 4 we first reformulate the noncommutative torus in a more geometrical form, and find a hidden symmetry. Next, in section 5, such a geometrical analysis will be used to formulate the unitary
projective representation of the group uniformizing arbitrary Riemann surfaces. In particular, we will focus on the problem of simultaneous covariantization of the differential operators representing the ${\rm sl}_2(\RR)$ generatoras and of their complex conjugate. This will lead to the covariantization of the conformal operators as given in (\ref{laTgrBAAA}).
We will also solve the problem of finding the explicit relation between the normal form of $\pi_1(\Sigma)$ and the uniformizing group $\Gamma$ of $\Sigma$. The extension to the nonabelian case is introduced in subsection \ref{nonabelext}. Section 6 is devoted to the properties of the gauge length as pseudo-distance and will show that it corresponds to a Poincar\'e area.

\vspace{.5cm}

\section{Differential representation of the conformal group}\label{passage}

In this section, after shortly reviewing the uniformization theorem for Riemann surfaces, we express the ${\rm PSL}_2(\RR)$ transformations, acting on the upper half-plane, in terms of the composition of three exponentiations of the generators of ${\rm sl}_2(\RR)$, represented by the differential operators
$\ell_k=z^{k+1}\partial_z$, $k=-1,0,1$. This will lead to consider two questions. The first is that such a prescription is not general, since the order of the composition depends on the specific M\"obius transformation. The other question is related to the problem of expressing the unitary operators ${\cal U}_k$ in the form of a unique exponential
\be
{\cal U}_k=\exp({\cal D}_k) \ .
\ee As we will see, this is the natural way to define the covariantization of the
conformal group. Such preliminary questions are solved in the next section, where we will derive the form of a M\"obius transformation in the form of a unique exponential, whose argument is a linear combination of the $\ell_k$'s.

\subsection{Uniformization and Liouville equation}\label{uniformization}

In the following, given a $2\times2$ matrix
$$
\mu=\left(\begin{array}{c}a\\ c
\end{array}\begin{array}{cc}b\\ d\end{array}\right) \ ,$$
we adopt the notation
$$
\mu z={az+b\over cz+d} \ .
$$
Let  ${\DD}$ be either the Riemann sphere
$\hat\CC=\CC\cup\{\infty\}$, the complex plane $\CC$, or the
upper half--plane
$$\HH=\{z\in\CC|\Im(z)>0\} \ .
$$ According to the uniformization
theorem, every Riemann surface $\Sigma$ is conformally
equivalent to ${\DD}/\Gamma$, where $\Gamma$ is a freely
acting discontinuous group of fractional transformations
preserving ${\DD}$.
Let $J_\HH$ be
the uniformizing map
$J_\HH:\HH\longrightarrow\Sigma$. It has the invariance property
$$ J_\HH(\gamma z)=J_\HH(z) \ ,
$$
$\gamma\in\Gamma$, where $\Gamma\subset{\rm PSL}_2(\RR)={\rm
SL}_2(\RR)/\{\pm I\}$  is a finitely generated
Fuchsian group. It acts on $\HH$ by the linear fractional
transformations
$$
\gamma z={az+b\over
cz+d}\in\HH \ ,\quad\qquad\gamma=\left(\begin{array}{c}a\\
c\end{array}\begin{array}{cc}b\\ d
\end{array}\right)\in\Gamma \ ,
$$
$z\in\HH$.
By the fixed point equation $\gamma z=z$, that is
$$
z_\pm={a-d\pm\sqrt{(a+d)^2-4}\over2c} \ , $$
it follows that there are three kinds of ${\rm
PSL}_2(\RR)$ matrices.

\vspace{0.4cm}

\begin{enumerate}
\item{{\it Elliptic:} $|{\rm tr}\,\gamma|<2$. Then $\gamma$ has
one fixed point on $\HH$ ($z_-=\overline z_+\notin\RR$) and
$\Sigma$ has a branched point $w_-=J_\HH(z_-)$.
The finite order of its stabilizer defines its index $n\in{\NN}\backslash\{0,1\}$. If $\Gamma$ contains elliptic
elements, then $\HH/\Gamma$ is an orbifold.}
\item{{\it Parabolic:} $|{\rm tr}\,\gamma|=2$. Then $z_-=z_+\in
\RR$ and the point $J_\HH(z_-)$ corresponds to a missing point of $\Sigma$,
{\it i.e.} a puncture.
The order of the
stabilizer is now infinite.}
\item{{\it Hyperbolic:} $|{\rm tr}\,\gamma|>2$. The fixed points are
distinct and lie on $\RR=\partial \HH$.
Such $\gamma$'s
correspond to handles of $\Sigma$ and
can be expressed in the form\footnote{Given $z$ and $\gamma
z$, $\lambda(\gamma)$ corresponds to the minimal distance between
them. This minimum is reached for $z$ lying on the geodesic
intersecting the real axis at $z_-$ and $z_+$.}
$${\gamma z
-z_+\over \gamma z-z_-}=e^\lambda{z-z_+\over z-z_-} \ ,
$$ $e^\lambda\in
\RR\backslash\{0,1\}$.}
\end{enumerate}

\vspace{0.4cm}

\noindent
The Poincar\'e metric on $\HH$ is the metric with scalar curvature
$-1$
\be
d{s}^2={|dz|^2\over(\Im(z))^2} \ .
\ee
This implies that the Liouville equation on $\Sigma$
\be
\partial_{\bar w}\partial_w\varphi={e^{\varphi}\over2} \ ,
\ee
has the unique solution
\be
e^{\varphi}={|{J_\HH^{-1}(w)}'|^2\over (\Im J_\HH^{-1}(w))^2} \ .
\ee
The basic property of the
Poincar\'e metric is that its isometry group ${\rm PSL}_2(\RR)$
coincides with the automorphism group of $\HH$. The group
$\Gamma$ is isomorphic to the fundamental group $\pi_1(\Sigma)$.

\vspace{0.4cm}

\noindent
If $\Gamma$ uniformizes a surface of genus $g$ with $n$ punctures and
$m$ elliptic points with indices $2\le n_1\le n_2\le\ldots\le
n_m<\infty$, then $\Gamma$ is generated by
$2g$ hyperbolic elements $\gamma_1,\ldots,\gamma_{2g}$, $m$
elliptic elements $E_1,\ldots,E_m$, and $n$ parabolic elements
$P_1,\ldots,P_n$ satisfying the relations
\begin{equation}
E_j^{n_j}=I \ ,\qquad
\Big(\prod_{l=1}^mE_l\Big)\Big(\prod_{k=1}^nP_k\Big)
\prod_{j=1}^g\left(\gamma_{2j-1}\gamma_{2j}{\gamma_{2j-1}^{-1}}
{\gamma_{2j}^{-1}}\right)=I \ .\label{3xpergamma}\end{equation}
The uniformizing group carries information both on
the topological and the complex structures of the Riemann surface.
It is easy to see that the number of parameters fixing the
generators of a Fuchsian group coincides with the dimension of
moduli space of complex structures of Riemann surfaces. For
example, in the case of compact Riemann surfaces of genus $g$, the
full set of generators depends on $6g$ real parameters which
reduce to $6g-3$ upon using (\ref{3xpergamma}). On the other hand,
(\ref{3xpergamma}) remains invariant under conjugation by an
element of ${\rm SL}_2(\RR)$ leading to $3g-3$ complex
parameters.

\subsection{Differential representation of ${\rm sl}_2(\RR)$}\label{drepalg}

Let $\ell_{n}$, $n=-1,0,1,$ be the generators of ${\rm sl}_2(\RR)$. Fix their normalization
by
$$
[\ell_m,\ell_n]=(n-m)\ell_{m+n}   \ ,$$
$n=-1,0,1$. Consider the representation $\ell_n\mapsto z^{n+1}\partial_z\in{\rm
End}(\CC[z,z^{-1}])$. Thus, we set
$$
\ell_n=z^{n+1}\partial_z \ .$$ Note that $[\ell_n,f]=z^{n+1}\partial_z
f$.
Each element of ${\rm PSL}_2(\RR)$ can
be expressed as the composition of a translation, dilatation and a
special conformal transformation. These ${\rm PSL}_2(\RR)$
transformations are
\begin{equation}
\exp({\lambda_{-1}\ell_{-1}})z=z+\lambda_{-1}\ ,\qquad
\exp({\lambda_0\ell_0})z=e^{\lambda_0}z \ ,\qquad
\exp({\lambda_1\ell_1})z={z\over 1-\lambda_1z} \ .
\label{viraasinistra}\end{equation} If $f(z)$ admits a convergent
series expansion, then
$$
\exp({\lambda_j\ell_j})f(z)=f(\exp({\lambda_j\ell_j})z) \ . $$ We note that
this action can be defined by considering the formal Taylor
theorem. This corresponds to equating the application of a formal
exponential of a formal multiple of $\partial_z$ with a formal
substitution operation. For example,
$$
\exp(\lambda_{-1}\ell_{-1})f(z)=f(z+\lambda_{-1}) \ . $$ In this case
$f(z)$ is an {\it arbitrary} series, including the formal ones. In particular, this
series may have the form $\sum_n a_n z^n$, with $n$ which may also
take complex values. Note that
the formal expansion of $\exp(\lambda_{-1}\ell_{-1})$, should not be confused with
the standard meaning of formal series expansion, referring to non-convergent expansions.
In particular, note that given an element $X$ of a Lie algebra $\mathfrak g$, equipped with a norm $||\cdot||$, $\exp(X)$ is well defined on all
$\CC$ whenever $||X||$ is finite.

\noindent
Recall that
$$\mu_j(\mu_k z)=(\mu_j\mu_k)(z) \ ,
$$ and
set $h_j(z)=\exp({\lambda_j\ell_j}) z$.
Since
\begin{equation}
\exp({\lambda_k\ell_k})\exp({\lambda_j\ell_j})f(z)=\exp({\lambda_k\ell_k})f(h_j(z))
=f(h_j(h_k(z))) \ ,
\label{actionofelle}\end{equation} it follows that
$\exp({\lambda_k\ell_k})\exp({\lambda_j\ell_j})$ acts in reverse order
with respect to the matrix representation. This implies that
the representation $V$ constructed in terms of the above
differential operators acts as
have
\begin{equation}
V(\mu)f(z)=f(\mu^{-1}z) \ ,\label{9iwiunbivs}\end{equation} so that,
since
$$
V(\mu)V(\nu)f(z)=f(\nu^{-1}\mu^{-1}z)=f((\mu
\nu)^{-1}z)=V(\mu\nu)f(z) \ ,$$ the homomorphism property is
preserved. This fixes the representative of $\mu\in{\rm
PSL}_2(\RR)$ to be
\begin{equation}
V(\mu)=\exp({\lambda_{-1}\ell_{-1}})\exp({\lambda_0\ell_0})\exp({\lambda_1\ell_1}) \ ,
\label{csi}\end{equation} where the $\lambda_k$'s are the ones
corresponding to the group element $\mu^{-1}$.

\subsection{Parameters of the representation}\label{matrix}

In order to derive the relation between the $\lambda_k$'s and
$\mu^{-1}$, we note that by (\ref{viraasinistra})
$$
\exp({\lambda_{-1}\ell_{-1}})\exp({\lambda_0\ell_0})\exp({\lambda_1\ell_1})z=
{z+\lambda_{-1}\over-\lambda_1z+e^{-\lambda_0}
-\lambda_{-1}\lambda_1}={Az+B\over Cz+D} \ . $$ Since $AD-BC=1$, we
have
\begin{equation}
D=\pm
e^{\lambda_0/2}\left(e^{-\lambda_0}-\lambda_{-1}\lambda_1\right) \ ,
\label{pepperepepeppe}\end{equation} so that
\begin{equation}
\lambda_{-1}={B\over A}\ ,\qquad e^{\lambda_0}={A^2}\ ,\qquad
\lambda_1=-{C\over A}\ ,
\label{duf}\end{equation}
and
$$
{Az+B\over Cz+D}=\exp\Big({{B\over A}\ell_{-1}}\Big)\exp(2\ln(A){\ell_0})\exp\Big({-{C\over
A}\ell_1}\Big)z \ .$$
\\
Since a global sign is irrelevant for ${\rm
PSL}_2(\RR)$ matrices, we can choose the $+$ sign on the right
hand side of (\ref{pepperepepeppe}), so that
\\
\begin{equation}
\left(\begin{array}{c}A\\ C\end{array}\begin{array}{cc}B\\
D\end{array}\right)=\left(\begin{array}{c}e^{\lambda_0/2}\\
-\lambda_1e^{\lambda_0/2}\end{array}\begin{array}{cc}\quad
\lambda_{-1}e^{\lambda_0/2}\\ \quad e^{-\lambda_0/2}-\lambda_{-1}
\lambda_1e^{\lambda_0/2}
\end{array}\right)  \ .
\label{staproviso}\end{equation}
\\ The above decomposition holds
when $A\neq0$. Group elements not attainable by the decomposition
(\ref{csi}) can be reached by alternative decompositions. A decomposition which holds when $D\neq0$ is the reversed
one
\be
\exp({\lambda_1\ell_1})\exp({\lambda_0\ell_0})\exp({\lambda_{-1}\ell_{-1}})z={
(e^{\lambda_0}-\lambda_{-1}\lambda_1)z+\lambda_{-1}\over
-\lambda_1 z+1}={Az+B\over Cz+D} \ .
\ee Actually,
\begin{equation}
A=\pm
e^{-\lambda_0/2}\left(e^{\lambda_0}-\lambda_{-1}\lambda_1\right) \ ,
\label{pepperepepeppe2}\end{equation} so that
$$
\lambda_{-1}={B\over D} \ ,\qquad e^{\lambda_0}=D^{-2} \ ,\qquad
\lambda_1=-{C\over D} \ , $$ and
$$
{Az+B\over Cz+D}=\exp\Big(-{C\over D}\ell_1\Big)\exp(-2\ln(D){\ell_0})\exp\Big({{B\over
D}\ell_{-1}}\Big)z \ . $$
\\
Choosing the $+$ sign on the right hand side of
(\ref{pepperepepeppe2}), we have \\
$$
\left(\begin{array}{c}A\\ C\end{array}\begin{array}{cc}B\\
D\end{array}\right)=\left(\begin{array}{c}e^{\lambda_0/2}-
\lambda_{-1}\lambda_1e^{-\lambda_0/2}\\\lambda_{-1}
e^{-\lambda_0/2}\end{array}\begin{array}{cc}\quad-\lambda_{-1}
e^{-\lambda_0/2}\\ \quad e^{-\lambda_0/2}\end{array}\right) \ .
$$ \\
Finally, if both $A$ and $D$ vanish, we can choose
\be
\exp({\lambda_{-1}\ell_{-1}})\exp({\lambda_{-1}^{-1}\ell_1})
\exp({\lambda_{-1}\ell_{-1}})z=-{\lambda_{-1}^2\over z} \ ,
\ee
so that
$$
B=-C^{-1}=\pm\lambda_{-1} \ ,
$$
where the sign ambiguity reflects the fact that this operator coincides with
the one associated to the inverse matrix.

\noindent
In the following we will need the operator version of
Eq.(\ref{9iwiunbivs}), that is
$$
V(\mu)f(z)V(\mu^{-1})=f(\mu^{-1}z) \ .
$$
Note that the
relation (\ref{3xpergamma}) is represented
by\footnote{We will use $\mu$ and $\nu$ to denote generic elements
of ${\rm PSL}_2(\RR)$, while $\beta$ and $\gamma$ will denote
elements of the uniformizing group $\Gamma\subset{\rm
PSL}_2(\RR)$.}
\begin{equation}
V\left[\prod_{j=1}^g\left(\gamma_{2j-1}\gamma_{2j}{\gamma_{2j-1}^{-1}}
{\gamma_{2j}^{-1}}\right)\right]=\prod_{j=1}^g\left(
V_{2j-1}V_{2j}V_{2j-1}^{-1}V_{2j}^{-1}\right)=I \ ,
\label{fwU}\end{equation} where we restricted to the hyperbolic
case and
\begin{equation}
V_k\equiv V(\gamma_k) \ .\label{csiA}\end{equation}

\section{Parameterization of ${\rm SL}_2(\RR)$ by the closed forms of the BCH formula}\label{BCHD}

We have seen that in order to express an element of ${\rm SL}_2(\RR)$ in the form
\be
\exp(\lambda_i\ell_i)\exp(\lambda_j\ell_j)\exp(\lambda_k\ell_k) \ ,
\label{siamointeressati}\ee
one has to distinguish three different cases, namely
\be
A\neq0 \ , \qquad D\neq0 \ , \qquad A=D=0 \ .
\label{threedistinguishedcases}\ee
On the other hand, to find the covariant form of the conformal group, we must express
(\ref{siamointeressati}) in the form
\be
\exp(\lambda_i\ell_i)\exp(\lambda_j\ell_j)\exp(\lambda_k\ell_k)=\exp(\mu_1\ell_1+\mu_2\ell_2+\mu_3\ell_3) \ ,
\label{theproblemmm}\ee
for any choice of $i$, $j$ and $k$.
Very recently, have been derived new closed forms for the Baker-Campbell-Hausdorff (BCH) formula that include, as a particular case,
the problem (\ref{theproblemmm}). Such a solution also solves the previous question, namely the one of writing
down a unique expression, in the form on the right hand side of (\ref{theproblemmm}), holding simultaneously for the three
distinguished cases (\ref{threedistinguishedcases}). The reason is that any group element can be always expressed in the form
on the right hand side of (\ref{theproblemmm}). Note that the unique exception would be if the trace of the corresponding matrix is $-2$
in the case when such a matrix is non-diagonalizable. On the other hand, we are interested in the linear fractional transformations where there is
no any problem. The expression on the right of (\ref{theproblemmm}) is quite general. This has a deep consequence since it means that the covariantization of the conformal group is {\it always definite}.
Let us further illustrate such a point, by considering $X$, $Y$ and $Z$, elements of an arbitrary Lie algebra $\mathfrak g$. Suppose that it has been found a finite linear combination $W$ of the generators
of  $\mathfrak g$ such that
\be
\exp(X)\exp(Y)\exp(Z)=\exp(W) \ .
\label{nearto}\ee
 For any suitable norm, the power expansion of $\exp(X)\exp(Y)\exp(Z)$ converges on all $\CC$, except in the points where one or more of the norms $||X||$, $||Y||$ and $||Z||$ are singular. It follows that if
 (\ref{nearto}) holds in a neighborhood of the identity, then it
should hold in a wider region with respect to the one related to the expansion of $\ln(\exp(X) \exp(Y)\exp(Z))$.
 Of course, this is related to the possible singularities of the norm of $W$. Remarkably, as observed in \cite{Matone:2015wxa}, this never happens for ${\rm PSL}_2(\CC)$.

\noindent The problem of finding the closed form of $W$ in (\ref{nearto}), has been recently considered in
\cite{Matone:2015wxa} where it has been introduced an algorithm that solves the BCH problem for a wide class of cases. In \cite{Matone:2015xaa} it has been shown that there are {\it 13 types} of commutator algebras admitting such closed forms of the BCH formula.
Furthermore, it turns out that the algorithm includes all the semisimple complex Lie algebras \cite{Matone:2015oca}, as they correspond to particular cases
of commutator algebras of {\it type 1c-i}, {\it type 4} and {\it type 5}. It turns out that ${\rm sl}_2(\RR)$ corresponds to a particular subtype of the {\it type} 4 commutator algebras.

\vspace{.4cm}

\noindent
Let us shortly show the main steps of the algorithm.
In \cite{Van-Brunt:2015ala}
Van-Brunt and Visser  obtained
a remarkable relation providing the closed form of the BCH formula in important cases. If $X,Y\in{\mathfrak g}$ satisfy the commutation relations
\begin{equation}
{}[X,Y]=uX+vY+cI \ ,
\label{beoftheform}\end{equation}
with $I$ a central element and $u$, $v$, $c$, complex parameters, then \cite{Van-Brunt:2015ala}
\begin{equation}
\exp(X) \exp(Y)= \exp({X+Y+f(u,v)[X,Y]}) \ ,
\label{bbbvvvII}\end{equation}
where $f(u,v)$ is the symmetric function
\be
f(u,v)={(u-v)e^{u+v}-(ue^u-ve^v)\over uv(e^u-e^v)} \ .
\ee
Such a result generalizes to a wider class of cases
by exploiting the associativity of the BCH formula and implementing the Jacobi identity \cite{Matone:2015wxa}. Consider the identity
\begin{equation}
\exp(X)\exp(Y) \exp(Z)= \Big(\exp(X)\exp({\alpha Y})\Big)\Big(\exp({\beta Y}) \exp(Z)\Big) \ ,
\label{decomposizione}\end{equation}
where
$$
\alpha+\beta=1 \ .
$$
If
\begin{equation}
[X,Y]=uX+vY+cI \ , \qquad [Y,Z]=wY+zZ+dI \ ,
\label{riecco}\end{equation}
then, by (\ref{bbbvvvII}),
\begin{equation}
{}\exp(X) \exp(\alpha Y) = \exp({\tilde X}) \ , \qquad \exp(\beta Y)\exp(Z) = \exp({\tilde Y}) \ ,
\label{associativity}\end{equation}
with
\begin{align}
\tilde X &:=g_{\alpha}(u,v)X+h_{\alpha}(u,v)Y+l_{\alpha}(u,v)cI \ , \cr
\tilde Y&:=h_{\beta}(z,w)Y+g_{\beta}(z,w)Z+l_{\beta}(z,w)dI \ ,
\label{colon1}\end{align}
where
\begin{align}
g_\alpha(u,v)&:=1+\alpha uf(\alpha u, v) \ , \cr
h_{\alpha}(u,v)&:=\alpha(1+vf(\alpha u, v)) \ , \cr
l_{\alpha}(u,v)&:=\alpha f(\alpha u, v) \ .
\end{align}
This solves the BCH problem since, by (\ref{beoftheform}), (\ref{bbbvvvII}), (\ref{decomposizione}) and (\ref{associativity}), it follows that
imposing
\begin{equation}
[\tilde X,\tilde Y]=\tilde u \tilde X+\tilde v \tilde Y+\tilde c I \ ,
\label{labella}\end{equation}
that fixes $\alpha$, $\tilde u$, $\tilde v$ and $\tilde c$, gives
\begin{equation}
\exp(X) \exp(Y) \exp(Z)= \exp(\tilde X) \exp(\tilde Y)=\exp({\tilde X+\tilde Y+f(\tilde u,\tilde v)[\tilde X,\tilde Y]}) \ .
\label{llasol}\end{equation}
Note that, consistently with the Jacobi identity
\be
[X,[Y,Z]]+[Y,[Z,X]]+[Z,[X,Y]]=0 \ ,
\label{lajacobbbi}\ee
$[X,Z]$ may contain also $Y$
\begin{equation}
[X,Z]=m X+nY+pZ+e I \ .
\label{richiamare}\end{equation}
The Jacobi identity
constrains $e, m,n$ and $p$ by a linear system.
Furthermore, note that setting $Y=\lambda_0Q$ and $\lambda_-:=\lambda_0\alpha$, $\lambda_+:=\lambda_0\beta$, Eq.(\ref{decomposizione})
implies, as a particular case,
\begin{equation}
\exp(X) \exp(Z)=\lim_{\lambda_0\to 0} \exp(X)\exp({\lambda_- Q}) \exp({\lambda_+ Q}) \exp(Z) \ ,
\label{decomposizionetre}\end{equation}
explicitly showing that the algorithm solves also the BCH problem for $\exp(X) \exp(Z)$,
in some of the cases when $[[X,Z],X]$ and $[[X,Z],Z]$ do not vanish.

\noindent In \cite{Matone:2015wxa}, as a particular case of the Virasoro algebra, it has been shown that the closed form of the BCH formula in the case of ${\rm sl}_2(\CC)$ is
\begin{align}
\exp({\lambda_{-1}\ell_{-1}})&\exp({\lambda_0\ell_0})
\exp({\lambda_1\ell_1})= \cr
&\exp\Big\{{\lambda_+-\lambda_-\over e^{-\lambda_-}-e^{-\lambda_+}}[\lambda_{-1} \ell_{-1}+(2-e^{-\lambda_+}-e^{-\lambda_-})\ell_0+\lambda_1 \ell_1]\Big\} \ ,
\label{lasoluzionee}\end{align}
where
\begin{equation}
e^{-\lambda_{\pm}}={1+e^{-\lambda_0}-\lambda_{-1}\lambda_1\pm \sqrt{(1+e^{-\lambda_0}-\lambda_{-1}\lambda_1)^2-4e^{-\lambda_0}}\over2} \ .
\end{equation}
Note that, as explained above, since the $\ell_k$'s are the differential representation of ${\rm sl}_2(\RR)$, it follows that the corresponding M\"obius transformations are
the ones associated to the inverse matrix. In fact, one may check that the correspondence between the $\lambda_k$'s and the matrix elements given
in Eqs.(\ref{pepperepepeppe})(\ref{duf}), is the same of the one in \cite{Matone:2015wxa} after the change
\begin{equation}
\left(\begin{array}{c}A\\ C\end{array}\begin{array}{cc}B\\
D\end{array}\right)\qquad \longrightarrow \qquad
\left(\begin{array}{c}D\\ -C\end{array}\begin{array}{cc}-B\\
A\end{array}\right) \ .
\label{portaa}\ee

\section{Setting the problem}\label{setting}

In the next sections we will construct the projective unitary
representation of the fundamental group $\pi_1(\Sigma)$ by means
of differential operators ${\cal U}_{k}$ acting on the Hilbert space $L^2(\HH)$.
This will lead to a relation for such operators which has the form
of (\ref{3xpergamma}) except that the identity on the right hand side is multiplied by a phase.
In the hyperbolic case we will have
\begin{equation}
\prod_{k=1}^g{\cal U}_{2k-1}\,{\cal U}_{2k}\,{\cal
U}_{2k-1}^{-1}\,{\cal U}_{2k}^{-1}=e^{2\pi i\theta}I \ .
\label{inonni}\end{equation}

\noindent
As a first step, here we consider the  unitarity problem.
In higher genus we will see the appearance of several new
structures. For example, a distinguished feature concerns the
combination of differential operators that one may use to
construct the unitary operators. While on the torus the
exponentials $e^{\partial_{x_1}}$ and $e^{\partial_{x_2}}$ both appear
separately, for $g>1$ the possible operators are restricted to a
specific combination of $\partial_1\equiv\partial_{x_1}$ and $\partial_2\equiv\partial_{x_2}$.

\noindent
In this section we will also reconsider the formulation of the noncommutative torus and, in particular, the way
the phase $e^{2\pi i\theta}$ in (\ref{noncommtoruspre}) is obtained. The aim is to learn from the torus as
much as possible, in order to reformulate the derivation in a way
which extracts the features of the construction without referring
to the specifics of the torus.

\noindent
The fact that the fundamental group in $g>1$ is nonabelian implies
that, in order to determine the phase in (\ref{inonni}), we cannot use the
reduced BCH formula
\begin{equation}
e^Ae^B=e^{[A,B]}e^Be^A \ ,\label{bch}\end{equation} which holds when
$A$ and $B$ commute with $[A,B]$. For $g=1$ the
associated differential generators commute, {\it i.e.}
$[\partial_1,\partial_2]=0$, so that it makes sense to use (\ref{bch})
to evaluate the phase coming from the (constant)
commutator of the covariant derivatives. For $g>1$, a computation
of the phase by means of the complete BCH formula would involve
quantities which are a covariantization of the already noncommuting
operators such as the generators of ${\rm sl}_2(\RR)$. However, this is not only a technical
difficulty, rather we actually still do not know which structures the covariantization of the generators of ${\rm sl}_2(\RR)$
may have.
As we said, reformulating the case of the torus
in a different language will suggest its natural
higher genus generalization, without using the complete BCH formula. In particular, it will shed lights on the covariantization procedure. As we will
see, the result is deeply related to the geometry of Riemann surfaces.
In particular, constructing the unitary operators that will projectively
represent $\pi_1(\Sigma)$, will bring us to a problem that can be
seen as the one of {\it simultaneous covariantization}.
Essentially, this is the problem of finding in higher genus the covariant version
\begin{equation}
{\cal O}_A=F{\cal O}F^{-1}, \label{ppoll}\end{equation} of a given
operator ${\cal O}$ in such a way that its adjoint has the form
${\cal O}_A^\dagger=F\tilde{\cal O}F^{-1}$, with $\tilde{\cal O}$
independent of the connection $A$ and $F$ a
functional of $A$.

\subsection{A first screening}\label{41}

Mimicking the case of $g=1$, where each one of the two operators
${\cal U}_1$ and ${\cal U}_2$ are expressed in terms of real
coordinates $x\equiv x_1$ and $y\equiv x_2$ respectively, one
expects that the building blocks for the solution to the quotient
conditions in $g>1$ have two possible forms, either
$$
\exp({{\cal L}_n-{\cal L}_n^{\dagger}}) \ ,
$$
or
$$
\exp({i({\cal L}_n+{\cal L}_n^{\dagger})}) \ ,
$$ where the ${\cal L}_n$'s are some {\it covariantized
${\rm sl}_2(\RR)$ operators} to be determined. Such operators should be the generalization to the case of the
three generators of ${\rm sl}_2(\RR)$ of the covariant
derivative.  Since in the case of the torus the relevant phase is expressed by
means of the commutator between covariant derivatives, and
considering that ${\cal L}_n$ contains
$\partial_z=\partial_x-i\partial_y$,
one may expect that,
in the case of higher genus Riemann surfaces, both $\exp({{\cal L}_n-{\cal
L}_n^{\dagger}})$ and $\exp({i({\cal L}_n+{\cal L}_n^{\dagger})})$
appear. However, while the exponentials $\exp({\partial_x})$ and
$\exp({\partial_y})$ generate translations, that map $\CC$, the universal covering of the torus, to itself, the operators $\exp({{\cal L}_n-{\cal
L}_n^{\dagger}})$ and  $\exp({i({\cal
L}_n+{\cal L}_n^{\dagger})})$ should generate ${\rm PSL}_2(\RR)$ transformations.
On the other hand, $\exp({i({\cal
L}_n+{\cal L}_n^{\dagger})})$ cannot generate real M\"obius transformations, so that
 we should discard $\exp({i({\cal
L}_n+{\cal L}_n^{\dagger})})$ and restrict to
  $\exp({{\cal L}_n-{\cal
L}_n^{\dagger}})$ only. This fact is strictly related to the
nonabelian nature of the group $\pi_1(\Sigma)$ which, in turn, is
related to the condition $\Im(z)>0$ defining $\HH$. The latter
reflects the fact that the translation operator along the
imaginary axis $\exp({\partial_y})$ alone does not belong to the automorphisms group
of $\HH$. Since among the translation operators
$\exp({\partial_x})$ and $\exp({\partial_y})$ only the former is allowed,
we see that comparing $\exp({{\cal L}_n-{\cal L}_n^{\dagger}})$ with
$\exp({\partial_x})=\exp({(\partial_z+\partial_{\bar z})/2})$ one should
expect that ${\cal L}_n^{\dagger}$ corresponds to $-\bar{\cal L}_n$.
We will see that a slightly modified version of this holds.

\noindent
Finding the ${\cal L}_n$'s is a problem closely related to the one
of deriving the central extension for the Fuchsian group without
using the BCH formula. Since the ${\cal L}_n$'s are covariant
versions of the three generators of ${\rm sl}_2(\RR)$, in general
the nested commutators
\begin{equation}
[{\cal
L}_{j_1},[{\cal L}_{j_2},[\cdots[{\cal L}_{j_{n-1}},{\cal
L}_{j_n}]]\cdots]] \ ,
\label{nobch}\end{equation}
should be difficult to treat, so that, apparently, one should use the complete version of the BCH formula. Nevertheless, since we
will succeed in finding the central extension of the uniformizing group, this implies that
the same result should be obtained by using the complete BCH
formula. Therefore, in spite of (\ref{nobch}), the
structure of the ${\rm sl}_2(\RR)$ implies a
simplification. In particular, determining the $c_{j_1\ldots
j_n}$ in
\begin{equation}
\sum_{l,m=-1}^1c_{j_1\ldots j_n} [{\cal L}_l,{\cal L}_m] =[{\cal
L}_{j_1},[{\cal L}_{j_2},[\cdots[{\cal L}_{j_{n-1}},{\cal
L}_{j_n}]]\cdots]] \ , \label{nobchb}\end{equation} should reveal a
considerable simplification of the complete BCH formula for the
case at hand.

\subsection{An alternative to the BCH formula on the
torus}\label{alternative}

Here we revisit the covariantization of the translation operators. These are naturally associated to the tessellation of the plane, so that
their covariantization leads to consider the algebra of the noncommutative torus. For sake of simplicity, we consider the covariantization of translations operators along
the Cartesian coordinates of $\RR^2$, denoted by $x_1$ and $x_2$. These operators are naturally associated to orthogonal tori.
The investigation will lead to the computation of the phase in Eq.(\ref{noncommtoruspre})
without using the BCH formula. This alternative will indicate
the way to covariantize the ${\rm PSL}_2(\RR)$ operators, therefore providing the extension of our analysis to $g>1$, where the direct use of the
BCH formula is, apparently, inextricable.

\noindent Let us consider the connection
$$
A=A_1 dx_1+A_2dx_2 \ ,
$$
and the operators
\begin{equation}
{\cal U}_k=\exp({\lambda_k(\partial_k+iA_k)}) \ ,
\label{oiedj}\end{equation}
$k=1,2$, where $\lambda_k\in\RR$.
Given two operators $A$ and $B$, and a function $f(B)$ satisfying
suitable properties, we have
$$
Af(B)A^{-1}=f(ABA^{-1}) \ . $$ We define the functions $F_k(x_1,x_2)$, $k=1,2$,
by
$$
{\cal U}_k=F_k\exp({\lambda_k\partial_k})F_k^{-1} \ ,
$$ that compared
with (\ref{oiedj}) yields
$$
(\partial_k+iA_k)F_k=0 \ .
$$ The solution of this equation is
$$
F_1(x_1,x_2)=\exp\Big(-i\int_{(x_1^0,x_2)}^{(x_1,x_2)} dx A_1(x,x_2)\Big) \ ,
$$
where the contour integral is with $x_2$ fixed. Similarly
$$
F_2(x_1,x_2)=\exp\Big(-i\int_{(x_1,x_2^0)}^{(x_1,x_2)} dx A_2(x_1,x)\Big) \ ,
$$
where the contour integral is with $x_1$ fixed. In the following we will use
the notation
$$
F_k=\exp\Big({-i\int_{x^0_k}^{x_k}da_kA_k}\Big) \ ,
$$ where in the integrand one
has $A_1(a_1,x_2)$ if $k=1$ and $A_2(x_1,a_2)$ if $k=2$.
There is an observation that simplifies considerably the construction. The key point is that since in the two contour integrals either $x_1$ or $x_2$ are fixed, so that either $dx_1=0$ or $dx_2=0$,
it follows that both the integrands, $da_1 A_1$ and $da_2 A_2$, can be replaced by the full connection $A=A_1dx_1+A_2 dx_2$. Therefore,
$$
F_k=\exp\Big({-i\int_{x^0_k}^{x_k}A}\Big) \ .$$
Since $\exp({\lambda_k\partial_k})$ is the translation operator, we have
\begin{align}
{\cal U}_k&=\exp\Big({-i\int_{x_k^0}^{x_k}A}\Big)\exp({\lambda_k\partial_k})
\exp\Big({i\int_{x_k^0}^{x_k}A}\Big) \cr
&=\exp\Big({i\int_{x_k}^{x_k+\lambda_k}
A}\Big)\exp({\lambda_k\partial_k}) \ .\label{thereasonforthis}\end{align}
This allows for a very geometric
derivation of the phase in (\ref{noncommtoruspre}).

\noindent
Since we are investigating the covariantization of $\partial_{x_1}$ and $\partial_{x_2}$, we consider tori
whose fundamental domain ${\cal F}$ is a rectangle. Denote by  $\lambda_1$ and $\lambda_2$ its base and height respectively.
By (\ref{thereasonforthis}) and
Stokes' theorem \\
$$
{\cal U}_1{\cal U}_2{\cal U}_1^{-1}{\cal U}_2^{-1}=$$
$$
\exp\left[i\int_{(x_1,x_2)}^{(x_1+\lambda_1,x_2)}A+
i\int_{(x_1+\lambda_1,x_2)}^{(x_1+\lambda_1,x_2+\lambda_2)}A
+i\int_{(x_1+\lambda_1,x_2+\lambda_2)}^{(x_1,x_2+\lambda_2)}A
+i\int_{(x_1,x_2+\lambda_2)}^{(x_1,x_2)}A\right]=
$$
\begin{equation}
\exp\left(i\oint_{\partial{\cal F}}A\right)=\exp\left(i\int_{\cal
F} F\right),\label{opijwJD}\end{equation} where $F$ is the curvature of $A$
\be
F=dA=(\partial_1A_2-\partial_2A_1)dx_1\wedge dx_2=F_{12}dx_1\wedge
dx_2 \ .
\ee
The above shows that the phase does {\it not} equal the
curvature $F$. The fact that our derivation and the one made with
the reduced BCH formula (\ref{bch}) coincide, is due to the fact that a possible solution is
given by a constant $F_{12}$. A possible choice to get a constant
phase is to set $F_{12}=2\pi\theta/\lambda_1\lambda_2$ that
 corresponds to
$$
A_1=-\pi{\theta\over\lambda_1\lambda_2} x_2 \ ,\qquad A_2=\pi{\theta
\over\lambda_1\lambda_2} x_1 \ .
$$
that by (\ref{bch}) and (\ref{oiedj}) gives
$$
\exp(\lambda_1(\partial_1+iA_1)) \exp(\lambda_2(\partial_2+iA_2))=
$$
\be
\exp(\lambda_1\lambda_2[\partial_1+iA_1,\partial_2+iA_2])\exp(\lambda_2(\partial_2+iA_2)) \exp(\lambda_1(\partial_1+iA_1)) \ ,
\ee
that is (\ref{noncommtoruspre}).
However, only in the case in which $F_{12}$
is constant does one have
$$
\int_{\cal F} F=\lambda_1\lambda_2F_{12} \ .
$$ Let us now show why
apparently the constant curvature connection is the unique
solution. Let us add the suffix $x_1x_2$ to ${\cal F}$ in order to
indicate its dependence on the base-point. Also, note that ${\cal
F}$ is univocally determined by $x_1$ and $x_2$. We need to show
that the integral of $F$ on ${\cal F}_{x_1x_2}$ is independent of
the point $(x_1,x_2)$. That is
\begin{equation}
\int_{{\cal F}_{x_1x_2}}F=\int_{{\cal F}_{x_1'x_2'}}F \ ,
\label{fjewhft}\end{equation} for any $(x_1',x_2')\in\RR^2$. Any
point in $\RR^2$ can be obtained by a translation
$$(x_1,x_2)\to(x_1',x_2')=\mu(x_1,x_2)\equiv (x_1+b_1,x_2+b_2) \ . $$
Let us apply the translation $\mu$ to the entire fundamental
domain and denote it by $\mu{\cal F}_{x_1x_2}$. Since ${\cal
F}_{x_1'x_2'}=\mu{\cal F}_{x_1x_2}$, we have
$$
\int_{{\cal F}_{x_1'x_2'}} F= \int_{{\cal F}_{x_1x_2}}\mu^\star F \ ,
$$
so that Eq.(\ref{fjewhft}) is
satisfied only if
$$
\int_{{\cal F}_{x_1x_2}}(\mu^\star F-F)=0 \ .
$$
This fixes $F$ to be a
constant two--form, up to a non--constant
contribution with vanishing surface integral. This implies a hidden invariance that we consider below.

\subsection{The hidden invariance}\label{distances}

Let us still consider tori with rectangular fundamental domain.
We now show that the above investigation also allows to find an
additional invariance of the operators
$$
{\cal U}_k=\exp({\lambda_k
(\partial_k+iA_k)}) \ ,
$$ under a suitable transformation of the
connection $A_k$. Such an invariance, which is not evident by
analyzing the ${\cal U}_k$'s expressed in the form
$\exp({\lambda_k(\partial_k+iA_k)})$, is a consequence of the fact
that $\exp({\lambda_k\partial_k})$ is a translation operator. By
(\ref{thereasonforthis}) we see that also ${\cal U}_k$ is a
translation operator
$$
{\cal U}_1f(x_1,x_2)=f(x_1+\lambda_1,x_2){\cal U}_1 \ , \qquad {\cal U}_2f(x_1,x_2)=f(x_1,x_2+\lambda_2){\cal U}_2 \ .
$$
Therefore, the ${\cal U}_k$'s have the invariance property
\begin{align}
h_k{\cal U}_kh_k^{-1}&=h_k\exp\Big({-i\int_{x_k^0}^{x_k}A}\Big)
\exp({\lambda_k\partial_k})\exp\Big({i\int_{x_k^0}^{x_k}A}\Big)h_k^{-1} \cr
&=\exp\Big({i\int_{x_k}^{x_k+\lambda_k} A}\Big)\exp({\lambda_k\partial_k})={\cal
U}_k \ ,
\end{align} whenever the $h_k$'s satisfy
\begin{equation}
h_1(x_1,x_2)h_1^{-1}(x_1+\lambda_1,x_2)=1 \ ,\qquad
h_2(x_1,x_2)h_2^{-1}(x_1,x_2+\lambda_2)=1 \ .
\label{conditionsh}\end{equation} To preserve
unitarity of ${\cal U}_k$, we require $|h_k|=1$. Thus,
the transformation of the connection
\begin{equation}
A_k \longrightarrow A_k+i\partial_k\ln h_k \ ,
\label{ofiwjfT3}\end{equation}
leaves ${\cal U}_k$ invariant.

\noindent In general (\ref{ofiwjfT3}) is not a gauge
transformation, as each component transforms according to a
different $h_k$. Consequently, under (\ref{ofiwjfT3}) the
curvature $F$ transforms to
$$
\tilde F=F-i\partial_1\partial_2\ln {h_1\over h_2}dx_1\wedge dx_2 \ ,
$$ while the phase (\ref{opijwJD}) remains
invariant. This transformation can be restricted to one component
of the connection only, that is we can choose either $h_1$ or $h_2$ to be a constant. Also, the
operators ${\cal U}_k$ are not gauge invariant unless the gauge
function is periodic up to a multiple of $2\pi$. That is, by (\ref{conditionsh}) it follows that ${\cal
U}_k$ is invariant under the gauge transformation
$$A\to A+d\chi \ ,
$$
if and only if
$$\chi(x_1+\lambda_1,x_2)=\chi(x_1,x_2)+2m\pi \ , \qquad \chi(x_1,x_2+\lambda_2)=\chi(x_1,x_2)+2m\pi \ ,
$$
$m,n\in\ZZ$. Of course, this is consistent with the fact that the transformation (\ref{ofiwjfT3}) is not a
gauge transformation unless $h_1=cnst\, h_2$.

\noindent
Above we saw how to remove the gauge connection $A$ from the covariant
derivative in ${\cal U}_k$.  The result is that the gauge connection
appears integrated along straight lines, that is geodesics with respect to the
flat metric, the right language to extend the construction to higher genus. Upon factorization, the
connection is acted on by the inverse of the differential operator
(integration). This shows a close relation between integrals of
the connection along geodesics, {\it i.e.}
$\int_{x_k}^{x_k+\lambda_k}A$, and covariantized operators.
More precisely, inverting (\ref{thereasonforthis}) yields
$$
\exp\Big({i\int_{x_k}^{x_k+\lambda_k}A}\Big)=\exp({\lambda_k(\partial_k+iA_k)})
\exp({-\lambda_k\partial_k}) \ .
$$

\section{Unitary projective representations of the
Fuchsian group}\label{differential}

\subsection{The problem of simultaneous covariantization}\label{problem}

In the following we will see that, as expected, in higher genus it is convenient to use complex coordinates $z$ and $\bar z$
rather than real ones $x$ and $y$. Some of the operators we
will work with have the structure
$$
\exp({\nabla_z-\nabla_z^\dagger}) \ ,
$$ which is unitary by construction.
The operators $\nabla_z$ and $\nabla_z^\dagger$ will be some
appropriate covariantization of $\partial_z$ and $\partial_{\bar z}$. Since
we will also consider the covariantization of the ${\rm sl}_2(\RR)$
differential operators, the noncommutativity of the latter will
naturally lead to the method described in the previous
subsections, rather than to the reduced BCH formula (\ref{bch}).

\noindent
The alternative derivation of the phase considered above may be
applied to other cases if one can set\footnote{The above cannot
correspond to $F(z,\bar z)\exp({\partial_z-\partial_{\bar
z}})F^{-1}(z,\bar z)$, as $\exp({\partial_z-\partial_{\bar z}})$ would
correspond to translations of $\Im(z)$ by an imaginary constant
and cannot be unitary. On the torus we can also use
$\exp({i(\nabla_z+\nabla_z^\dagger))}$. However, as seen in subsection
\ref{41}, the nonabelian nature of $\pi_1(\Sigma)$ forbids the use
of $i$.}
\begin{equation}
\exp({\nabla_z-\nabla_z^\dagger})=F\exp({\partial_z+\partial_{\bar
z}})F^{-1} \ ,
\label{osiqjbisseee}\end{equation} for some suitable
function $F(z,\bar z)$. The fact that we need such simultaneous
covariantization, $i.e.$, that we need to express both the covariant
derivative and its adjoint as standard derivatives conjugate by the
{\it same} function, has already been used in the case of the torus. In
that case we considered the action of
unitary operators on the $F_k$ rather than
the reduced BCH formula. It is clear that if we
had, for example,
$\nabla_z\propto F\partial_z F^{-1}$ and
$\nabla_z^\dagger\propto F^{-1}\partial_{\bar z} F$, then it would
not be possible to extend the above method to $g>1$. In that case
Eq.(\ref{osiqjbisseee}) would not hold, and it would not be
possible to get the phase on the right hand side of (\ref{inonni}) by using the action of differential operators
$$
\exp({\partial_z+\partial_{\bar z}})F^{-1}(z,\bar z)=F^{-1}(z+1,\bar
z+1)\exp({\partial_z+\partial_{\bar z}}) \ .
$$
As an example, one can consider differential operators acting on
automorphic functions on the upper half--plane, or, what is
equivalent, covariant operators acting on sections of line bundles
on a Riemann surface $\Sigma$\footnote{To simplify notation, we will use
the same symbol $z$ to denote a coordinate both on $\HH$ and
$\Sigma$.}. These operators can be seen as a sort of covariantization of
$\partial_z$ and $\partial_{\bar z}$. More precisely, consider the
metric tensor $g_{z\bar z}$, so that $ds^2=2g_{z\bar z}dz d\bar z$.
The covariant derivative acting on $K^\lambda$, where $K$ is the
canonical line bundle on $\Sigma$, is
$$
\nabla_z^\lambda: K^\lambda\longrightarrow K^{\lambda+1} \ ,
$$ where
\begin{equation}
\nabla_z^\lambda\psi=g_{z\bar z}^\lambda{\partial_z}g_{z\bar
z}^{-\lambda}\psi=(\partial_z-\lambda\partial_z\ln g_{z\bar
z})\psi \ .
\label{ouoahx}\end{equation} Formally, one can consider
this as a suitable covariantization of $\partial_z$. The scalar product on
$K^\lambda$ is
$$
\langle\phi|\psi\rangle=\int_\Sigma d\nu\sqrt g\left(g^{z\bar
z}\right)^\lambda\bar\phi\psi \ ,
$$ where
$$
d\nu(z)={i\over2}dz\wedge d\bar z \ .
$$ It follows that
the adjoint of $\nabla_z^\lambda$
$$
(\nabla_z^\lambda)^\dagger: K^{\lambda+1}\longrightarrow
K^\lambda \ ,
$$ is
\begin{equation}
(\nabla_z^\lambda)^\dagger\psi=-g^{z\bar z}\partial_{\bar z}\psi \ .
\label{slkjP}\end{equation} In the literature there are also other
examples of covariant differential operators on higher genus
Riemann surfaces. For example one can consider covariant operators
acting on sections of $K^\mu\otimes\bar K^\nu$, and then take the
dual space to be sections of $K^\rho\otimes\bar K^\sigma$ for some
$\rho$ and $\sigma$. In particular, for a suitable choice of the
weights $\mu,\nu,\rho$ and $\sigma$, one can obtain covariant
operators of the form $g_{z\bar z}^\lambda\partial_zg_{z\bar
z}^{-\lambda}$, with adjoints $-g_{z\bar
z}^{-\lambda}\partial_{\bar z}g_{z\bar z}^\lambda$ which, in a
certain sense, exhibit more symmetry than (\ref{ouoahx}) and
(\ref{slkjP}). Nevertheless, also in this case we have
$\partial_z$ conjugate by $g_{z\bar z}^\lambda$ and
$g_{z\bar z}^{-\lambda}$, whereas $\partial_{\bar z}$ is
conjugate in the reverse order, that is by $g_{z\bar z}^{-\lambda}$ and $g_{z\bar
z}^\lambda$.
Thus, an apparently unavoidable feature of covariant operators is
that they never admit the simultaneous covariantization
\begin{equation}
\nabla_z-\nabla_z^\dagger=F(z,\bar z)(\partial_z+\partial_{\bar
z})F^{-1}(z,\bar z) \ ,\label{oidhqw}\end{equation} for some $F$. On
the other hand, we have just seen that we need precisely the
property that, given a covariant differential operator (in particular, an
${\rm sl}_2(\RR)$ differential operator) and its adjoint, both of
them should be expressed as a non covariant operator conjugate by
$F$ and $F^{-1}$ in the same order.

\noindent
Let us further illustrate this point. In evaluating the adjoint
operator, one performs an integration by parts and simultaneously
takes the complex conjugate. Complex conjugation is the crucial
point. In fact, if we now construct covariantized operators by
conjugating them by $F$ and $F^{-1}$, with $F$ a phase, then
complex conjugation corresponds to the inversion of $F$. Thus,
the unique solution is to choose
$$
\bar F=F^{-1}\ ,
$$ and define
\begin{equation}
\nabla_z=F\partial_zF^{-1} \ .
\label{oihjYh}\end{equation} Note that
the condition $|F|=1$ essentially follows also by requiring the
unitarity of the operator in (\ref{osiqjbisseee}). Let us consider
the scalar product
$$
\langle\phi|\psi\rangle=\int_\HH d\nu\bar\phi\psi \ ,
$$ integrating
by parts, by (\ref{oihjYh}) we have
$$
\langle\phi|\nabla\psi\rangle=\int_\HH d\nu\bar\phi
F\partial_zF^{-1}\psi=-\int_\HH d\nu\overline{F\partial_{\bar
z}(F^{-1}\phi})\psi=\langle\nabla_z^\dagger\phi|\psi\rangle \ ,
$$
showing that the adjoint $\nabla_z^\dagger$ is constructed by
conjugating $\partial_{\bar z}$ in the same way as
$-\partial_z$ is conjugate to get $\nabla_z$
$$
\nabla_z^\dagger=-F\partial_{\bar z}F^{-1} \ . $$ Therefore, we found
the way of covariantizing the derivatives as in (\ref{oidhqw}). Later on
we will apply the above method to the case of the ${\rm
sl}_2(\RR)$ differential operators.

\noindent
The above integration domain is $\HH$ rather than the
Riemann surface itself, as was the case previously. Our equations
will be defined on the upper half--plane, as we are interested in
constructing a projective unitary representation of the group
$\pi_1(\Sigma)$ by means of operators acting on $L^2(\HH)$. In
particular, the action of our operators will not be restricted to
automorphic forms, which is the case when the equations are to be
projected onto the Riemann surface. In this respect we now show
that trying to perform a similar trick in the case of scalar
products defined on a Riemann surface would lead to imaginary
powers of the metric. Denoting by $\psi^{(\mu,\sigma)}$ a section
of $K^\mu\otimes\bar K^{\sigma}$, we see that
\begin{equation}
\int_{\Sigma}\overline{\phi^{(\rho,i\kappa)}}g_{z\bar z}^{i\kappa}
\partial_zg_{z\bar z}^{-i\kappa}\psi^{(i\kappa,1-\rho)}=-\int_{\Sigma}
\overline{g_{z\bar z}^{i\kappa}\partial_{\bar z}\left(g_{z\bar
z}^{-i\kappa}
\phi^{(\rho,i\kappa)}\right)}\psi^{(i\kappa,1-\rho)} \ ,
\label{oihPG}\end{equation} where $\rho,\kappa$ are real
numbers.\footnote{The construction can be generalized to the case
in which the weight $\rho$ is a complex number. The only
difference consists in replacing $\psi^{(i\kappa,1-\rho)}$ by
$\psi^{(i\kappa,1-\bar\rho)}$. One may also consider replacing
$i\lambda$ by a complex number $\mu$, however in such a case
$g^\mu_{z\bar z}$ would not be a phase and simultaneous covariantization
would not be possible.} The integrand in (\ref{oihPG}) is a
$(1,1)$--form, and the action of the derivatives is covariant,
that is, they act on $0$--differentials (e.g. $\partial_zg_{z\bar
z}^{-i\kappa}\psi^{(i\kappa,1-\rho)}$). Also, the complex
conjugate of a $(\mu,\nu)$ differential is a $(\bar\nu,\bar\mu)$
differential.\footnote{Differentials have been studied in the
literature with real
\cite{Bonora:1989wk}\cite{Bonora:1989zm}
and complex weights
\cite{Voronov:1990pp}.} Eq.(\ref{oihPG}) implies that the adjoint of
\begin{equation}
\nabla_z=g_{z\bar z}^{i\kappa}\partial_z g_{z\bar z}^{-i\kappa} \ ,
\label{oficvjw}\end{equation} is
\begin{equation}
\nabla_z^\dagger=-g_{z\bar z}^{i\kappa}\partial_{\bar z}g_{z\bar
z}^{-i\kappa} \ .
\label{soihHy}\end{equation} As a consequence of the
fact that both $\nabla_z$ and $\nabla_z^\dagger$ are obtained by
conjugating $\partial_z$ and $-\partial_{\bar z}$ by
$g_{z\bar z}^{i\kappa}$ and $g_{z\bar z}^{-i\kappa}$, we have that
a function of any linear combination of $\nabla_z$ and of its adjoint
has the property
$$
f(a\nabla_z+b\nabla_z^\dagger)=g_{z\bar
z}^{i\kappa}f(a\partial_z+b
\partial_{\bar z})g_{z\bar z}^{-i\kappa} \ .
$$
The appearance of the phase ties together several mathematical
aspects which have a physical meaning.
In particular, considers the operators
(\ref{oficvjw}) and (\ref{soihHy}) on the upper half--plane
endowed with the Poincar\'e metric
$$
ds^2=y^{-2}|dz|^2=2g_{z\bar z}|dz|^2=e^{\varphi}|dz|^2 \ ,
$$ then one
would obtain $\nabla_z=y^{-2i\kappa}\partial_z y^{2i\kappa}$ and
$\nabla_z^\dagger=-y^{-2i\kappa}\partial_{\bar z}y^{2i\kappa}$.
In this respect, it
is interesting to observe that the Poincar\'e Laplacian
$$
\Delta=-4y^2\partial_{\bar z}\partial_z \ ,
$$ satisfies the equation
$$
\Delta y^{{1\over2}+i\kappa}=\lambda_\kappa
y^{{1\over2}+i\kappa} \ ,
$$ where the eigenvalues are
\begin{equation}
\lambda_\kappa={1\over 4}+\kappa^2 \ .
\label{pwoKDg}\end{equation}
The problem of simultaneous covariantization led us to introduce imaginary
powers of the metric. In turn, the associated Laplacian has
eigenfuctions corresponding to complex powers of such a metric. On
the other hand, the appearance of this complex power lies at the
heart of the mass gap $1/4$ in (\ref{pwoKDg}) which never appears
in flat spaces such as in the case of the torus. Thus, there is a strict relationship
between the noncommutativity of the fundamental group $\pi_1(\Sigma)$, the
structures derived from imposing the simultaneous covariantization, and the
structure of the Laplacian eigenvalues themselves. The mass gap
$1/4$ is in fact a sort of regularization induced by the negative
curvature (which in turn is related to the nonabelian nature of
$\pi_1(\Sigma)$). In this context, it is worth mentioning that in
\cite{Callan:1989em}
curvature brings the infrared and ultraviolet behavior of QCD
under analytic control without any conflict with gauge invariance.

\subsection{The unitary covariantized operators}\label{unitarygaugedop}

Set
$$e_n(z)=z^{n+1} \ ,$$
$n=-1,0,1$,
and define the operators
$$
L_n=e^{1/2}_n\partial_ze^{1/2}_n=e_n^{-1/2}\ell_ne^{1/2}_n=
e_n(\partial_z+\partial_z\ln e^{1/2}_n) \ ,
$$ that is
$$
L_1=z^2\partial_z+z=\ell_1+z \ ,\qquad
L_0=z\partial_z+{1\over2}=\ell_0+{1\over2} \ ,\qquad
L_{-1}=\partial_z=\ell_{-1}\ .
$$ This deformation of the $\ell_n$
has no effect on the algebra, that is
$$
[L_m,L_n]=(n-m)L_{m+n} \ .
$$ Furthermore
$$[L_n,f]=z^{n+1}\partial_zf \ .$$ Let us define the {\it covariantized}
operators
$$
{\cal L}_n^{(F)}=FL_n F^{-1}=Fe^{1/2}_n\partial_ze^{1/2}_n
F^{-1}=e_n[\partial_z+\partial_z\ln(e^{1/2}_n F^{-1})] \ , $$ where
$F(z,\bar z)$ is an arbitrary function of unitary modulo
$$
|F|=1 \ . $$ Since the ${\cal L}_n^{(F)}$ and $L_n$ differ by a
conjugation, it follows that the ${\cal L}_n^{(F)}$ satisfy the same algebra of the $L_n$
$$
[{\cal L}_m^{(F)},{\cal L}_n^{(F)}]=(n-m){\cal L}_{m+n}^{(F)} \ .
$$
Introducing the scalar
product on $L^2(\HH)$
$$
\langle\phi|\psi\rangle=\int_\HH d\nu\bar\phi\psi \ , $$ we see that
an integration by parts leads to
$$
\langle\phi|{\cal L}_n^{(F)}\psi\rangle=-\int_\HH d\nu\overline{F
\overline{e^{1/2}_n}\partial_{\bar
z}\left(\overline{e^{1/2}_n}F^{-1}\phi\right)}\psi=\langle{\cal
L}_n^{(F)\dagger}\phi|\psi\rangle \ ,$$ that is the adjoint of ${\cal
L}_n^{(F)}$
\begin{equation}
{\cal L}_n^{(F)\dagger}=-F\overline{e^{1/2}_n}\partial_{\bar
z}\overline {e^{1/2}_n}F^{-1}=-\bar{\cal L}_n^{(F^{-1})} \ .
\label{abbastanzabasilare}\end{equation} Comparing
$$
{\cal L}_n^{(F)^\dagger}=(Fe_n^{1/2}\partial_ze_n^{1/2}
F^{-1})^{\dagger}=F{\overline
{e_n^{1/2}}}\partial_z^{\dagger}{\overline{e_n^{1/2}}}F^{-1} \ , $$
with (\ref{abbastanzabasilare}) one obtains
$$
\ell_n^{\dagger}=-\bar e_n^{-1}\bar\ell_n\bar e_n \ . $$ In the case
$n=-1$ we have $\partial_z^{\dagger}=-\partial_{\bar z}$, so that,
since $\partial_z=(\partial_x-i\partial_y)/2$, we see that the
construction reproduces the usual adjoint operation in the case of
$\partial_x$ and $\partial_y$.

\noindent
The basic property of the adjoint ${\cal L}_n^{(F)\dagger}$ is
that it is obtained by conjugating $-\bar L_n$ with $F$ and $F^{-1}$. This means that $L_n$ and $-\bar L_n$ are covariantized
in the same way. This solves the aforementioned conjugation problem. Actually, the operator
$$
\Lambda_n^{(F)}={\cal L}_n^{(F)}-{\cal L}_n^{(F)\dagger}={\cal
L}_n^{(F)}+\bar{\cal L}_n^{(F^{-1})} \ , $$ is the sum of $L_n$ and
$\bar L_n$ covariantized by means of the same conjugation
$$
\Lambda_n^{(F)}=F(L_n+\bar L_n)F^{-1} \ , $$ so that
$$
\exp({\Lambda_n^{(F)}})=F\exp({L_n+\bar L_n})F^{-1} \ . $$ Since
$\Lambda_n^{(F)\dagger}=-\Lambda_n^{(F)}$, we formally have
\begin{equation}
\exp({\Lambda_n^{(F)}})\exp({\Lambda_n^{(F)\dagger}})=I=
\exp({\Lambda_n^{(F)\dagger}})\exp({\Lambda_n^{(F)}}) \ .
\label{zx8}\end{equation} A rigorous proof of unitarity goes as
follows. Set
$$
g_n={|e_n(w)|\over|e_n(z)|} \ , $$ where\footnote{The coefficients
$a_n,b_n,c_n$ and $d_n$ are given by (\ref{staproviso}) with $\lambda_n=1$
and $\lambda_{m\neq n}=0$.}
$w=e^{\ell_n}ze^{-\ell_n}=(a_nz+b_n)/(c_nz+d_n)$, and note that
\begin{align}
\exp({\Lambda_n^{(F)}})&=F(z,\bar
z)|e_n(z)|^{-1}\exp({\ell_n+\bar\ell_n})|e_n(z)|F^{-1} (z,\bar
z) \cr
&=F(z,\bar z)g_n F^{-1}(w,\bar w)\exp({\ell_n+\bar\ell_n}) \ .
\end{align}
{}By
$$
[\ell_n,\exp({\ell_n})]z=0 \ , $$ it follows that
$$\partial_z(\exp({\ell_n}) z) \exp({-\ell_n})={1\over z^{n+1}} \exp({\ell_n}) z^{n+1}\exp({-\ell_n}) \ ,
$$
that is
$\partial_zw={e_n(w)/e_n(z)}$, so that $$g_n/|\partial_z
w|^2=1/g_n \ . $$  We
then see that the operators $e^{\Lambda_n^{(F)}}$ are unitary
\begin{eqnarray}
\langle\phi|e^{\Lambda_n^{(F)}}\psi\rangle&=&\int_\HH{d\nu(w)
\over\left|{\partial_zw}\right|^2}\bar\phi(z,\bar z)
F(z,\bar z) g_nF^{-1}(w,\bar w)\psi(w,\bar w)\nonumber\\
&=&\int_\HH d\nu(w)\overline{F(w,\bar w)g_n^{-1} F^{-1}(z,\bar z)
\phi(z,\bar z)}\psi(w,\bar w)\nonumber
\\ &=&\langle e^{-\Lambda_n^{(F)}}\phi|\psi\rangle \ ,\nonumber
\end{eqnarray}
where in the first equality we used the fact that ${\rm PSL}_2(\RR)$ is the automorphism group of $\HH$.

\subsection{Selecting the Fuchsian generators}

We now digress on the possible realizations of the fundamental
relation (\ref{3xpergamma}) for a Fuchsian group in the hyperbolic
case. Being differential operators, the $V_k$ in (\ref{csiA}),
have the property of acting in the reverse order with respect to
the matrix product. This aspect raises a subtlety in considering
the relationship between $\pi_1(\Sigma)$ and the uniformizing
group $\Gamma$. Namely, let us consider the normal form for
$\Sigma$. This is a polygon whose symbol is
\begin{equation}
a_1b_1a_1^{-1}b_1^{-1}\ldots a_gb_ga_g^{-1}b_g^{-1}=I \ ,
\label{iVcappaD}\end{equation} where $\{a_k,b_k\}$ is a basis for
$\pi_1(\Sigma)$. Cutting the surface along these cycles one
obtains a simply connected domain whose vertices are connected by
elements of the covering group. If one considers this domain as
sitting on the upper half--plane, then one can consider it as a
fundamental domain with the transformations connecting the
vertices given by elements of $\Gamma$. Let us order the vertices
of the polygon in the counterclockwise direction and denote them
by $z=z_0,z_1,\ldots z_{4g-1},z_{4g}=z$. We denote this
fundamental domain for $\Gamma$ by
\begin{equation}
{\cal F}_z[\Gamma]=\{z=z_0,z_1,z_2,\ldots,z_{4g-1},z_{4g}=z\} \ ,
\label{fundo}\end{equation} with the vertices joined by geodesics.
Note that since the geodesics are univocally determined it follows
that the $4g$--gon fundamental domain itself is univocally
determined by the action of the Fuchsian generators on the
base-point $z$.

\noindent
A consequence of (\ref{iVcappaD}) is that the elements of $\Gamma$
satisfy a similar fundamental relation. Among these, the one we
wrote for the $\gamma_k$ in (\ref{3xpergamma}), is the canonical
one, that is the one in which the generators appear in the
sequence $\gamma_{2j-1}\gamma_{2j}{\gamma_{2j-1}^{-1}}
{\gamma_{2j}^{-1}}$. This version is obtained using the
identification
$$
z_1=\gamma_1 z \ ,\quad z_2=\gamma_1\gamma_2 z \ , \quad z_3=\gamma_1
\gamma_2\gamma_1^{-1} z \ , \quad\ldots,\quad z_{4g}=
\prod_{j=1}^g\left(\gamma_{2j-1}\gamma_{2j}{\gamma_{2j-1}^{-1}}
{\gamma_{2j}^{-1}}\right)z=z \ . $$ Other representations can be found
$e.g.$ by Dehn twisting.\footnote{These correspond to in general
non simultaneous conjugation of $\Gamma$'s generators by suitable
strings of the generators themselves (see for example
\cite{FarkasKra}\cite{Novikov}).} Note that these transformations
leave $\Gamma$ invariant and should not be confused with the ones
obtained by conjugating $\Gamma$ in\footnote{While the conjugation
in ${\rm PSL}_2(\RR)$ can be used to fix three real parameters,
so that the number of real independent moduli reduces to $3\times
\sharp$ generators $-3$ (due to the fundamental relation) $-3$ (due to
conjugation)=$6g-6$, the Dehn twists correspond to a discrete set
of transformations whose existence implies the nontrivial orbifold
structure of the moduli space of Riemann surfaces.} ${\rm
PSL}_2(\RR)$. Anyway, the usual representation for the
fundamental relation satisfied by the generators of $\Gamma$ given
in (\ref{3xpergamma}) does not fit with the aim of our
construction. Actually, what we essentially need is to provide a
central extension of the Fuchsian group. In particular, we are
looking for operators provinding a projective representation of
$\Gamma$, such that the fundamental relation is modified by a
phase. To discuss this aspect we first need to introduce the
Fuchsian matrices $\beta_k$ defined by
\begin{equation}
z_k=\beta_kz_{k-1}\equiv{\beta_{k_{11}}z_{k-1}+\beta_{k_{12}}\over
\beta_{k_{21}}z_{k-1}+\beta_{k_{22}}} \ .\label{woirkk}\end{equation}
Since $z_{4g}=z$, we have that the $\beta_k$ satisfy the
fundamental relation
\begin{equation}
\beta_{4g}\beta_{4g-1}\ldots\beta_1=I \ .
\label{woirkk2}\end{equation} The associated operators providing a
differential representation of $\Gamma$ are
\begin{equation}
T_k\equiv T(\beta_k)=\exp({\lambda_{-1}^{(k)}(L_{-1}+\bar
L_{-1})})\exp({\lambda_0^{(k)}(L_0+\bar L_0)})
\exp({\lambda_1^{(k)}(L_1+\bar L_1)}) \ , \label{iVcappa}\end{equation}
where $\lambda_{-1}^{(k)}$, $\lambda_0^{(k)}$ and
$\lambda_1^{(k)}$ are defined in such a way that
\begin{equation}
T_kf(z,\bar z)T_k^{-1}=f(\beta_k^{-1}z,\beta_k^{-1}\bar z) \ .
\label{tresettescacchibiliardobriscolachitarra}\end{equation} The
characterizing property of the generators $\beta_k$ for $\Gamma$,
is that (\ref{woirkk}) allows us to associate $T_k$ to the geodesic
connecting $z_{k-1}$ and $z_k$. This is an essential point because
it will allow to obtain the phase of the central extension of
$\Gamma$ in terms of an integral whose contour coincides with the
fundamental domain. In particular, the covariantization of the $T_k$ will
be performed by multiplying the $T_k$ on the left by the abelian Wilson
line associated to the geodesic connecting $z_{k-1}$ and $z_k$.
For this reason we will call these covariantized versions of the $T_k$
abelian Wilson--Fuchs operators. Let us note that these Wilson lines
on the upper half--plane correspond to Wilson loops on the Riemann
surface.

\noindent
In order to derive the relationships between the $\gamma_k$ and
$\beta_k$ we compare
$$
z_1=\gamma_1 z \ ,\qquad z_2=\gamma_1\gamma_2 z \ ,\qquad
z_3=\gamma_1\gamma_2\gamma_1^{-1}z \ ,\qquad z_4=\gamma_1
\gamma_2\gamma_1^{-1}\gamma_2^{-1}z \ ,\ldots \ ,
$$
with
$$
z_1=\beta_1 z \ ,\qquad z_2=\beta_2\beta_1z \ ,\qquad
z_3=\beta_3\beta_2\beta_1z \ ,\qquad
z_4=\beta_4\beta_3\beta_2\beta_1z \ , \ldots \ , $$ to obtain
$$
\beta_1=\gamma_1 \ ,\quad\beta_2=\gamma_1\gamma_2\gamma_1^{-1} \ ,\quad
\beta_3=\gamma_1\gamma_2\gamma_1^{-1}
\gamma_2^{-1}\gamma_1^{-1} \ ,\quad\beta_4=\gamma_1\gamma_2\gamma_1^{-1}
\gamma_2^{-1}\gamma_1\gamma_2^{-1}\gamma_1^{-1} \ ,\ldots \ . $$ One can
check that the above relationships extend to
\begin{equation}
\beta_k=\rho_k\gamma_{\sigma_k}^{\epsilon_k}\rho_k^{-1} \ ,
\label{oieeU}\end{equation} $k\in[1,4g]$, where
$$
\rho_k=\prod_{j=1}^{k-1}\gamma_{\sigma_j}^{\epsilon_j} \ , $$
$\rho_1\equiv I$, and
$$
\epsilon_k=(-1)^{[(k-1)/2]} \ ,\qquad
\sigma_k=2\left[{2k+3-2[k/2]\over4}\right]+2[k/2]-k \ , $$ with
$[\cdot]$ denoting the integer part. Note that
$$
\epsilon_{4k}=\epsilon_{4k-1}=-1 \ ,\qquad \epsilon_{4k-2}=
\epsilon_{4k-3}=1 \ ,
$$
and
$$
\sigma_{4k}=\sigma_{4k-2}=2k \ ,\qquad
\sigma_{4k-1}=\sigma_{4k-3}=2k-1 \ ,
$$
$k\in[1,g]$. Eq.(\ref{oieeU}) implies
\begin{equation}
\beta_{4k-j}=\beta_{4k-j-1}\beta_{4k-j-2}^{-1}\beta_{4k-j-1}^{-1} \ ,
\label{oieeUBISSEvisisis}\end{equation} where $j=0,1$, and
$k\in[1,g]$. We can use these relationships between the $\beta_k$
to select $2g$ elements, say $\beta_{4k-3}$, $\beta_{4k-2}$,
$k\in[1,g]$ which can be seen as a complete set of generators for
$\Gamma$. In particular, by (\ref{oieeUBISSEvisisis}) we can
express $\beta_{4k-1}$, $\beta_{4k}$, $k\in[1,g]$ in the form
$$
\beta_{4k-1}=\beta_{4k-2}\beta_{4k-3}^{-1}\beta_{4k-2}^{-1} \ ,
$$
and
$$
\beta_{4k}=\beta_{4k-2}\beta_{4k-3}^{-1}\beta_{4k-2}^{-1}
\beta_{4k-3}^{-1}\beta_{4k-2} \ .
$$

\subsection{The Wilson-Fuchs operators}\label{nerbuson}

In (\ref{csiA}) we introduced the operators $V_k$ which are
defined in terms of the $\ell_n$. However, in order to construct
unitary operators, we should use the $L_n$'s rather than the
$\ell_n$'s. Nevertheless, since the algebras of the $\ell_n$'s and
$L_n$'s coincide, we have that the commutation properties between
the $V_k$, and therefore fundamental relation (\ref{fwU}), would
remain invariant if the $\ell_n$ in $V_k$ are replaced by the
$L_n$'s. Similarly the $T_k\equiv T(\beta_k)$ would satisfy the same
fundamental relation under the replacement $L_n\to\ell_n$, $\bar
L_n\to\bar\ell_n$.

\noindent
Let us now consider the operators $T_k$. Since their action is in
the reverse order with respect to the one of the matrix product,
by (\ref{woirkk2}) and
(\ref{tresettescacchibiliardobriscolachitarra}) we have
$$T_{4g}\ldots T_1 f(z,\bar
z)T_1^{-1}\ldots T_{4g}^{-1}=f(\beta_1^{-1}\ldots\beta_{4g}^{-1}
z,\beta_1\ldots\beta_{4g} \bar z)=f(z,\bar z) \ , $$ that
is\footnote{We observe that a possible phase on the right hand
side of (\ref{thetees}) is excluded by construction.}
\begin{equation}
T_{4g}T_{4g-1}\ldots T_1=I \ .\label{thetees}\end{equation}

\noindent
Let $A$ be a $U(1)$ connection. We set
\begin{equation}
d_A(z,w)=\int_z^wA \ ,\label{defellecazio}\end{equation} where the
contour of integration is the Poincar\'e geodesic connecting $z$
and $w$. Recall that Poincar\'e geodesics are semi--circles
centered on the real axis. Semi--circles through $\infty$
correspond to straight lines parallel to the imaginary axis. In
the following we will mainly be interested in the case in which
$w$ is dependent on $z$, in particular we will consider the
function
$$
d_A(z,\mu z)=\int_z^{\mu z}A \ , $$ where
$$
\mu z={az+b\over cz +d} \ ,
$$
$\mu\in\Gamma$.
Let $b$ be an arbitrary real number. Consider the Wilson loop
\be
W_{\beta_k}=\exp\Big({ib\int_z^{\beta_k^{-1}z}A}\Big) \ ,
\ee
and
define the
Wilson--Fuchs operators
\begin{equation}
{\cal U}_k=W_{\beta_k}T_k \ ,
\label{loperatoree}\end{equation} where the $T_k$ have been
defined in (\ref{iVcappa}). Each ${\cal U}_k$ defines the function
$F_k(z,\bar z)$ solution of the equation
\begin{equation}
F_kT_kF^{-1}_k={\cal U}_k \ . \label{starrr}\end{equation} Since by
(\ref{tresettescacchibiliardobriscolachitarra}) we have
$$
T_kF_k^{-1}(z,\bar z)=F_k^{-1}(\beta_k^{-1}z,\beta_k^{-1}\bar
z)T_k \ , $$ it follows that Eq.(\ref{starrr}) is equivalent to
\begin{equation}
F_k(\beta_k^{-1}z,\beta_k^{-1}\bar z)=\exp({-ibd_A(z,\beta_k^{-1}
z)})F_k(z,\bar z) \ .\label{loperatoreeeqBISSA}\end{equation} We also
note that since
$$
F_k\exp({L_n+\bar L_n})F_k^{-1}=\exp({F_k(L_n+\bar L_n)F_k^{-1}}) \ ,
$$
it follows by (\ref{starrr}) that
\begin{equation}
{\cal U}_k=F_kT_kF_k^{-1}=\exp({\lambda_{-1}^{(k)}\Lambda_{-1,k}})
\exp({\lambda_0^{(k)}\Lambda_{0,k}})\exp({\lambda_1^{(k)}\Lambda_{1,k}}) \ ,
\label{gliucappa}\end{equation} where
$$
\Lambda_{n,k}\equiv\Lambda_{n,k}^{(F_k)}=F_k(L_n+\bar
L_n)F^{-1}_k \ . $$
In particular,
$$
\exp({\Lambda_{n,k}})=F_k\exp({L_n+\bar L_n})F_k^{-1} \ . $$ Eqs.(\ref{zx8})
and (\ref{gliucappa}) show that the ${\cal U}_k$'s are unitary
operators
$$
{\cal U}_k{\cal U}_k^\dagger=I={\cal U}_k^\dagger{\cal U}_k \ . $$
Note that by (\ref{tresettescacchibiliardobriscolachitarra}) we
have
$$
{\cal U}_k^\dagger={\cal
U}_k^{-1}=T_k^{-1}\exp\Big({-ib\int_z^{\beta_k^{-1}z}A}\Big)=
\exp\Big({-ib\int_{\beta_kz}^zA}\Big)T_k^{-1} \ . $$

\subsection{Covariantizing the generators of the uniformizing group}

\noindent
We now have all the ingredients to define the covariantization of the generators of the uniformizing group.
We have seen how to covariantize a given operator of ${\rm sl}_2(\RR)$. The problem now is the following.
Let us write $T_k$ in the form
\begin{equation}
T_k=\exp({D_k}) \ .
\label{abbastanzabasilaredue}\end{equation}
Knowing $D_k$ allows one to define the covariant operator in the natural way, that is
\begin{equation}
{\cal D}_k:=F_kD_kF_k^{-1} \ ,
\label{thecov}\end{equation}
so that, by (\ref{gliucappa}),
\begin{equation}
{\cal U}_k= \exp({{\cal D}_k}) \ .
\label{anchequesta}\end{equation}
On the other hand, by
 (\ref{iVcappa}) we have
\begin{equation}
\exp({D_k})=\exp({\lambda_{-1}^{(k)}(L_{-1}+\bar
L_{-1})})\exp({\lambda_0^{(k)}(L_0+\bar L_0)})
\exp({\lambda_1^{(k)}(L_1+\bar L_1)}) \ .
\label{iVcappabisse}\end{equation}
It follows that in order to find $D_k$, and therefore ${\cal D}_k$, one needs to solve the BCH problem (\ref{iVcappabisse}). On the other hand, as
reviewed in section \ref{BCHD}, the problem (\ref{iVcappabisse}) belongs to the class of the new closed forms for the BCH formula derived in \cite{Matone:2015wxa}.
In particular, by (\ref{lasoluzionee}), (\ref{gliucappa}), (\ref{anchequesta}), and the fact that, for any, the $k$ $\Lambda_{j,k}$'s satisfy the ${\rm sl}_2(\RR)$ commutation relations,
with the same normalization of the ones satisfied by the $\ell_n$'s (and $L_n$'s),
it follows that
\be
{\cal D}_k=
{\lambda_+^{(k)}-\lambda_-^{(k)}\over e^{-\lambda_-^{(k)}}-e^{-\lambda_+^{(k)}}}\big[\lambda_{-1}^{(k)} \Lambda_{-1,k}+(2-e^{-\lambda_+^{(k)}}-e^{-\lambda_-^{(k)}})\Lambda_{0,k}+\lambda_1^{(k)} \Lambda_{1,k}\big] \ ,
\label{laTgr}\ee
where
\begin{equation}
e^{-\lambda_{\pm}^{(k)}}={1+e^{-\lambda_0^{(k)}}-\lambda_{-1}^{(k)}\lambda_1^{(k)}\pm \sqrt{\big(1+e^{-\lambda_0^{(k)}}-\lambda_{-1}^{(k)}\lambda_1^{(k)}\big)^2-4e^{-\lambda_0^{(k)}}}\over2} \ .
\end{equation}
Let us explicitly rewrite ${\cal D}_k$ in terms of differentials operators. We have
$$
{\cal D}_k=
{\lambda_+^{(k)}-\lambda_-^{(k)}\over e^{-\lambda_-^{(k)}}-e^{-\lambda_+^{(k)}}}\times
$$
\be
\times F_k\big[\lambda_{-1}^{(k)}(\partial_z+\partial_{\bar z})+(2-e^{-\lambda_+^{(k)}}-e^{-\lambda_-^{(k)}})
(z\partial_z+\bar z\partial_{\bar z}+1)+\lambda_1^{(k)}(z^2\partial_z+{\bar z}^2\partial_{\bar z}+z+\bar z)\big]F_k^{-1} \ .
\label{laTgrB}\ee
This is a basic result. Starting with the analogy with the case of the torus, we considered several issues, such as the problem of the simultaneous covariantization, that
led to consider various differential representations of ${\rm sl}_2(\RR)$, and then the problem of unitarity. This culminated with the use of the recent results on the BCH formula, that, finally,
implied (\ref{laTgrB}).

\subsection{Computing the phase}\label{introducing}

Let us consider the
following string of operators
$$
{\cal U}_1^{-1}\ldots{\cal U}_{4g}^{-1}
=\exp\Big({ib\int_z^{\beta_1 z}A}\Big) T_1^{-1}
\exp\Big({ib\int_z^{\beta_2z}A}\Big) T_2^{-1}\ldots \exp\Big({ib\int_z^{\beta_{4g
}z}A}\Big) T_{4g}^{-1} \ . $$ Moving the operators $T^{-1}_k$ on the right
we obtain
$$
{\cal U}_1^{-1}\ldots{\cal U}_{4g}^{-1}=\exp\Big({ib
\int_z^{\beta_1z} A+ib\int_{\beta_1z}^{\beta_2\beta_1z}A+\ldots
+ib\int_{\beta_{4g-1}\ldots\beta_1z}^{\beta_{4g}\ldots
\beta_1z}A}\Big)T_1^{-1}\ldots T_{4g}^{-1} \ , $$ that by
(\ref{thetees}) reads
\begin{equation}
{\cal U}_{4g}\ldots{\cal U}_1
=\exp\Big({-ib\oint_{\partial{\cal F}_z[\Gamma]}A}\Big) \ ,
\label{abbiamox1}\end{equation} where ${\cal F}_z[\Gamma]$ is the
fundamental domain (\ref{fundo}).

\noindent
Until now the construction concerned an arbitrary $U(1)$ connection $A$. We now consider the condition on $A$ in order to provide a unitary projective representation of the
central extension of $\Gamma$, the integral $\oint_{\partial{\cal
F}_z[\Gamma]}A$ should be $z$--independent. Let us first apply Stokes' theorem\footnote{Recall that our
choice of generators of the Fuchsian group corresponds to the
boundary $\partial{\cal F}_z$ being counterclockwise oriented.}
\begin{equation}
\oint_{\partial{\cal F}_z[\Gamma]}A=\int_{{\cal F}_z[\Gamma]} dA \ .
\label{jhvdwd}\end{equation} An arbitrary transformation of a
point $z$ in $\HH$ can be expressed as $z\to z'=\mu z$ for some
$\mu\in{\rm PSL}_2(\RR)$. The fundamental domain with base-point
$\mu z$ reads
$$
{\cal F}_{\mu z}[\Gamma]=\{\mu z,\beta_1\mu z,\beta_2\beta_1\mu
z,\beta_3\beta_2\beta_1\mu z,\ldots\} \ . $$ Therefore,
$z$--independence implies
\begin{equation}
\int_{{\cal F}_{\mu z}[\Gamma]}dA=\int_{{\cal
F}_{z}[\Gamma]}dA \ .\label{mueffezetaAQESBprima}\end{equation} We
now consider the fundamental domain obtained from ${\cal
F}_z[\Gamma]$ under the $\mu$ map. Since by definition the sides
of ${\cal F}_z[\Gamma]$ are geodesics and these transform to
geodesics under the action of ${\rm PSL}_2(\RR)$, it follows that
the image of a M\"obius transformation acting on the entire domain
${\cal F}_{z}[\Gamma]$, is uniquely fixed by the M\"obius
transformed vertices, that is
$$
\mu{\cal F}_z[\Gamma]=\{\mu z,\mu\beta_1 z,\mu\beta_2\beta_1 z,
\mu\beta_3\beta_2\beta_1 z,\ldots\} \ . $$ The domains ${\cal F}_{\mu
z}[\Gamma]$ and $\mu{\cal F}_{z}[\Gamma]$ coincide up to a
conjugation of $\Gamma$ by $\mu$, namely
\begin{equation}
{\cal F}_{\mu z}[\Gamma]=\mu{\cal F}_z[\mu^{-1}\Gamma\mu] \ .
\label{mueffezetaAQES}\end{equation} Since the representation
cannot depend on the concrete choice of the fundamental domain, we
should check that the connection $A$ satisfies
\begin{equation}
\int_{\mu{\cal F}_z[\mu^{-1}\Gamma\mu]}dA=\int_{\mu{\cal
F}_z[\Gamma]}dA \ .\label{mueffezetaAQESB}\end{equation} Thus, from
Eqs.(\ref{mueffezetaAQESBprima})(\ref{mueffezetaAQES}) and
(\ref{mueffezetaAQESB}) we have
$$
\int_{{\cal F}_{z}[\Gamma]}dA=\int_{\mu{\cal
F}_z[\Gamma]}dA=\int_{{\cal F}_z[\Gamma]}\mu^\star dA \ , $$ that is
$$
\int_{{\cal F}_{z}[\Gamma]}(dA-\mu^\star dA)=0 \ .
$$
This imply that $dA$ must be ${\rm PSL}_2(\RR)$--invariant.
It is well known
that, up to an overall constant factor, the Poincar\'e form
$$
\omega_P=2 g_{z\bar z} d\nu=e^\varphi{i\over2} dz\wedge d\bar z \ ,
\qquad e^\varphi=y^{-2} \ , $$ is the unique ${\rm
PSL}_2(\RR)$--invariant (1,1)--form. We choose the
proportionality factor to be 1, that is
\begin{equation}
F=dA=2i(\partial_{\bar z}A_z-\partial_z A_{\bar z})d\nu=\omega_P \ ,
\label{ofiI}\end{equation} that, up to gauge transformations, has
solution
$$
A_z=A_{\bar z}={1\over2y} \ , $$ so that
\begin{equation}
A=A_zdz+A_{\bar z}d\bar z={dx\over y} \ .\label{Aconn}\end{equation}
We will call $A$ the {\it Poincar\'e connection}. It is worth
recalling that the Poincar\'e metric
$$
ds^2=y^{-2}|dz|^2=2g_{z\bar z}|dz|^2 \ , $$ has constant negative
curvature
$$
R=-g^{z\bar z}\partial_z\partial_{\bar z}\ln\, g_{z\bar z}=-1 \ , $$
that is, $\varphi$ satisfies the Liouville equation
$$
\partial_{\bar z}\partial_z\varphi={e^\varphi\over2} \ . $$
This implies the following properties for $A$
$$
A_z=A_{\bar
z}={1\over2}e^{\varphi/2}=-{i\over2}\partial_z\varphi={i\over2}\partial_{\bar
z}\varphi={1\over\sqrt 2}(\partial_z\partial_{\bar
z}\varphi)^{1/2} \ , $$ and
$$
F=dA=4A_zA_{\bar z}d\nu \ . $$ {}From
$\sqrt{g}R=-{1\over2}e^{\varphi}$ and the Gauss--Bonnet theorem we
have
\begin{equation}
\int_{{\cal F}_z[\Gamma]}\omega_P=-2\pi\chi(\Sigma) \ ,
\label{gauss}\end{equation} where $\chi(\Sigma)=2-2g$ is the Euler
characteristic of $\Sigma$. By (\ref{jhvdwd})(\ref{ofiI}) and
(\ref{gauss}), we have
$$
\oint_{\partial{\cal F}_z[\Gamma]}A=-2\pi\chi(\Sigma) \ , $$ and
(\ref{abbiamox1}) finally becomes
\begin{equation}
{\cal U}_{4g}\ldots{\cal U}_1=\exp({2\pi
ib\chi(\Sigma)}) \ .\label{perfect}\end{equation}

\subsection{Nonabelian extension}\label{nonabelext}

Up to now we considered the case in which the connection is
Abelian. However, it is easy to extend our construction to the
nonabelian case in which the gauge group $U(1)$ is replaced by
$U(N)$. The operators ${\cal U}_k$ now become the path--ordered
exponentials
$$
{\cal U}_k=P\exp\Big({ib\int^{\beta_k^{-1} z}_z A}\Big) T_k \ , $$ where the $T_k$
are the same as before, times the $U(N)$ identity matrix.
Eq.(\ref{abbiamox1}) is replaced by
\begin{equation}
{\cal U}_{4g}\ldots{\cal U}_1=P
\exp\Big({-ib\oint_{\partial{\cal F}_z[\Gamma]}A}\Big) \ .
\label{abbiamox1nonabeliana}\end{equation} Given an integral along
a closed contour $\sigma_z$ with base-point $z$, the path--ordered
exponentials for a connection $A$ and its gauge transform
$A^U=U^{-1}AU+U^{-1}dU$ are related by \cite{THOMPSON}
\begin{align}
P\exp &\Big({i\oint_{\sigma_z}A}\Big)=U(z)P\exp\Big({i\oint_{\sigma_z}A^U}\Big)U^{-1}(z) = \cr
&U(z)P\left[\exp\left({i\oint_{\sigma_z}d\sigma^\mu\int_0^1ds
s\sigma^\nu
U^{-1}(s\sigma)F_{\nu\mu}(s\sigma)U(s\sigma)}\right)\right]U^{-1}(z) \ .
\end{align}
This implies that the only possibility to get a
coordinate--independent phase is for the curvature (1,1)--form
$F=dA+[A,A]/2$ to be the identity matrix in the gauge indices
times a (1,1)--form $\eta$, that is
$$
F=\eta I \ . $$ It follows that
\begin{equation}
P\exp\Big({-ib\oint_{\partial{\cal F}_z[\Gamma]}A}\Big)=\exp\Big({-ib\int_{{\cal
F}_z[\Gamma]}F}\Big) \ .\label{boh}\end{equation} This is only a necessary
condition for coordinate--independence. Reasoning as in the
Abelian case, one concludes that $\eta$ should be proportional to
the Poincar\'e (1,1)--form, that is
\begin{equation}
\eta=k\omega_P \ . \label{olpikfc}\end{equation} In order to fix the
constant $k$, we first consider the vector bundle ${\cal E}$ on
which the connection $A$ is defined. Taking the gauge group $U(N)$
in the fundamental representation, the dimension $N$ of the vector
fiber is called the {\it rank} of ${\cal E}$. The {\it degree} is
the integral of the first Chern class\footnote{Our convention for
$A$ differs from the one in the mathematical literature by a
factor $i$.}
$$
M={\rm deg}\,({\cal E})={1\over2\pi}{\rm tr}\,\int_{{\cal F}}F \ ,
$$ where, to simplify notation we used ${\cal F}$
instead of ${\cal F}_z[\Gamma]$. Since in our case the trace gives
a factor $N$, we have
\begin{equation}
\int_{{\cal F}}F=2\pi\mu({\cal E})I \ ,
\label{funita}\end{equation} where
$$
\mu({\cal E})={{\rm deg}({\cal E})\over{\rm rank}({\cal
E})}={M\over N} \ . $$ Thus, by (\ref{abbiamox1nonabeliana}) and
(\ref{boh}) we have
\begin{equation}
{\cal U}_{4g}\ldots{\cal U}_1=\exp\Big({-2\pi ib\mu({\cal
E})}\Big)I \ .\label{abbiamox1nonabelianaduino}\end{equation} Finally,
we observe that by (\ref{gauss}) and (\ref{funita}) it follows
that the constant in (\ref{olpikfc}) is $k=-{\mu({\cal
E})/\chi(\Sigma)}$, that is
$$
F=-{\mu({\cal E})\over\chi(\Sigma)}\omega_PI \ .
$$

\section{The gauge length}\label{GaugeHochschild}

By (\ref{Aconn}) it follows that Eq.(\ref{defellecazio}) becomes
$$
d_A(z,w)=\int^{w}_{z}{dx\over y} \ , $$ where, we recall,
the contour integral is along the Poincar\'e geodesic connecting
$z$ and $w$. Let us denote by $x_0$ the center of this geodesic and $\rho$ its radius. In polar coordinates we have $z-x_0=\rho
e^{i\alpha_z}$, $w-x_0=\rho e^{i\alpha_w}$ and $dx/y=\rho
d\cos\alpha/\rho\sin\alpha=-d\alpha$, so that
\begin{equation}
d_A(z,w)=-\int_{\alpha_z}^{\alpha_w}d\alpha=\alpha_z-\alpha_w \ .
\label{logaritmo}\end{equation} Since
$$
e^{i\alpha_{zw}}={e^{i\alpha_z}-e^{-i\alpha_w}\over
e^{i\alpha_w}-e^{-i\alpha_z}}={z-\bar w\over w-\bar z} \ ,
$$ where $\alpha_{zw}\equiv\alpha_z-\alpha_w$, we have
\begin{equation}
d_A(z,w)=\alpha_{zw}=-i\ln\left({z-\bar w\over w-\bar z}\right).
\label{owidhiI}\end{equation} Note that in terms of $z$ and $w$,
we have
$$
x_0={1\over2}{|w|^2-|z|^2\over\Re(w-z)} \ , $$ which shows the
dependence of $\alpha_z$ (and $\alpha_w$) on $z$ and $w$. Also
note that both $\alpha_z$ and $\alpha_w$ range between $0$ and
$\pi$, with the extremes corresponding to points on the extended real axis
$\RR\cup\{\infty\}=\partial\HH$.

\subsection{The gauge length as pseudo--distance}\label{pseudodistance}

We now show that
\begin{equation}
\ell_A(z,w)=|d_A(z,w)| \ ,
\label{basilare}\end{equation}
 is in fact a pseudo--distance that we call {\it gauge length}. The
symmetry property follows from the antisymmetry of $d_A(z,w)$
while the triangle inequality
\begin{equation}
\ell_A(z_1,z_2)+\ell_A(z_2,z_3)\geq\ell_A(z_1,z_3) \ ,
\label{schwarz}\end{equation} follows from the fact that
$\ell_A(z,w)=|\alpha_{zw}|$ or, equivalently, from the observation
that $\ell_A(z_1,z_2)+\ell_A(z_2,z_3)-\ell_A(z_1,z_3)$ is the Poincar\'e
area of the geodesic triangle with vertices $z_1,z_2$ and $z_3$.
Since
$$
\ell_A(z,w)=0    \quad {\rm iff} \quad\Re(z)=\Re(w) \ , $$
it follows that $\ell_A(z,w)$ cannot be a distance.

\noindent
Note that
Eq.(\ref{schwarz}), seen as an inequality involving angles, is
similar to the one satisfied by the angles of triangles in
Euclidean geometry. While in Euclidean geometry the Schwarz
inequality is satisfied both by the angles and by the lenghts of
the sides of triangles, in the case of hyperbolic geometry, the
gauge length coincides with the angles themselves.

Another property of this pseudo--distance is that, as we said, it
has $\pi$ as upper bound corresponding to the case in which the
two points are on the real axis, so that
$$
\ell_A(z,w)<\pi \ , \quad \forall (z,w)\in\HH^2 \ . $$

\subsection{The gauge length as Poincar\'e area}\label{gaulenpoinc}

The gauge length has some interesting properties which are worth
mentioning. For example, while the geodesic distance between a
point in $\HH$ and one on the real axis measured with respect to
the Poincar\'e metric is divergent, the corresponding gauge
distance is finite. As a consequence, measuring the gauge distance
between one point on a Riemann surface and a puncture on it gives
a finite result. In particular, the greatest gauge length between
two points is $\pi$, which is the gauge distance between two
punctures. Also, the lower bound for $d_A(z,w)$ is $0$, which
corresponds to the case in which two points have the same real
part.

\noindent
We saw that since $dA=e^{\varphi}d\nu$ is the infinitesimal Poincar\'e area, Stokes' theorem and Gauss--Bonnet formula give
$\oint_{\partial{\cal F}}A=-2\pi\chi(\Sigma)$. The Stokes formula
is also useful to describe the gauge length of a {\it single}
geodesic as Poincar\'e area. In this respect recall that the Poincar\'e area of a hyperbolic triangle of angles
$\alpha,\beta$ and $\gamma$, is
$\pi-\alpha-\beta-\gamma$. Then consider the geodesic triangle
${\cal D}$ corresponding to the geodesic joining $z$ and $w$,
together with the two geodesics connecting $z$ and $w$ to the
point at imaginary infinity, which is a cusp so that $\gamma=0$. The latter two geodesics correspond
to straight lines parallel to the imaginary axis, thus they have
vanishing gauge length. Then, by Stokes' theorem
$$
\ell_A(z,w)=\Big|\int_z^w A\Big|=\int_{\partial{\cal D}}A=\int_{{\cal D}} dA=\pi-\alpha-\beta \ ,
$$  giving the relation
$$\alpha_{zw}=\pi-\alpha-\beta \ , $$
that can be directly verified. Therefore, the gauge length has in fact
properties which are related to those of an area function.

\noindent
An interesting property of the gauge length concerns its
transformation properties under ${\rm PSL}_2(\RR)$ M\"obius
transformations. We have
\begin{equation}
d_A\left(\mu z,\mu w\right)=d_A(z,w)+{i\over2}
\ln{\bar\mu_z\mu_w\over\mu_z\bar\mu_w} \ ,
\label{ioeuIpo}\end{equation} where
$$
\mu x\equiv{ax+b\over cx+d} \ , $$ and
$$
\mu_x\equiv\partial_x\mu x={1\over (cx+d)^2} \ . $$ On the other hand
$$
{i\over2}\ln{\bar\mu_z\mu_w\over\mu_z\bar\mu_w}
=d_A(0,cz+d)+d_A(cw+d,0) \ , $$ showing that the M\"obius
transformation of the integration limits corresponds to adding the
gauge lengths between $0$ and $cz+d$ and between $cw+d$ and $0$
$$
\int_{\mu z}^{\mu
w}A=\int_z^wA+\int_0^{\mu_z^{-1/2}}A+\int_{\mu_w^{-1/2}}^0A \ . $$ In
this respect it is worth noticing that if $c>0$ then the points
$\mu_z^{-1/2}=(cz+d)\in\HH$ and $\mu_w^{-1/2}=(cw+d)\in\HH$ are
hyperbolic transformations of $z$ and $w$ respectively, that is $\mu_z^{-1/2}=\nu z$,
$\mu_w^{-1/2}=\nu w$ where
$$
\nu=\left(\begin{array}{c}c^{1/2}\\ 0\end{array}\begin{array}{cc}
bc^{1/2}\\ c^{-1/2}\end{array}\right) \ . $$ Denote by $z_k$, $k=1,\ldots,
n$, $n\geq3$, the vertices of an $n$--gon with angles $\alpha_1,\ldots,\alpha_n$. Its gauge length is given by
$$
\ell^{(n)}_A(\{z_k\})=\sum_{k=1}^n|\alpha_{z_kz_{k+1}}|=\pi(n-2)-\sum_{k=1}^n
\alpha_k \ , $$ where $z_{n+1}\equiv z_1$. Eq.(\ref{ioeuIpo}) implies that
the gauge length of an $n$--gon is ${\rm PSL}_2(\CC)$--invariant,
that is
\begin{equation}
\ell^{(n)}_A(\{\mu z_k\})=\ell^{(n)}_A(\{z_k\}) \ .
\label{opqjwdd}\end{equation} This invariance can be also seen by
observing that $\ell^{(n)}_A$ can be expressed in terms of cross
ratios. In particular, for the geodesic triangle we have
$$
\ell_A^{(3)}=|\ln (z_1,z_2,\bar z_2,\bar z_3)(z_2,z_3,\bar z_2,\bar
z_1)| \ , $$ where
$$
(x_1,x_2,x_3,x_4)={x_1-x_3\over x_1-x_4}{x_2-x_4\over x_2-x_3} \ .
$$ The invariance (\ref{opqjwdd}) is then a
consequence of the ${\rm PSL}_2(\CC)$--invariance
$$
(\mu x_1,\mu x_2,\mu x_3,\mu x_4)=(x_1,x_2,x_3,x_4)\ ,
$$ which together with the fact that any cyclic
permutation corresponds to an involution
$$
(x_4,x_1,x_2,x_3)={(x_1,x_2,x_3,x_4)\over (x_1,x_2,x_3,x_4)-1} \ ,
$$ constitute the main properties of the cross
ratio.

\subsection{M\"obius transformations as gauge
transformations}\label{mobiusgauge}

While on one hand Eq.(\ref{jhvdwd}) implies the equality
$$
\oint_{\partial{\cal F}}A=\int_{\cal F}\omega_P \ , $$ on the other
hand the Poincar\'e metric is ${\rm PSL}_2(\RR)$--invariant while
$A$ is not. As a consequence the variation of $A$ under a ${\rm
PSL}_2(\RR)$ transformation can only be a total derivative. This
is in fact the case, as
\begin{equation}
A(\mu z,\mu\bar z)=A(z,\bar
z)-i\partial_z\ln(cz+d)dz+i\partial_{\bar z}\ln(c\bar z+d)d\bar z \ .
\label{pokwI}\end{equation} Since $cz+d$ has no zeroes on $\HH$,
it follows that $\ln(cz+d)$ is well defined, so that the
inhomogeneous term in (\ref{pokwI}) can be expressed as an external
derivative on $\HH$, that is
$$
A(\mu z,\mu\bar z)=A(z,\bar z)+d\ln(\mu_z/\bar\mu_z)^{i\over2} \ ,
$$ and
$$
\int_z^w A(\mu x,\mu\bar x)=\int_z^wA(x,\bar x)+{i\over2}
\ln{\bar\mu_z\mu_w\over\mu_z\bar\mu_w}=\int_{\mu z}^{\mu w}
A(x,\bar x) \ . $$ Therefore, the isometry group of the Poincar\'e
metric, which in turn coincides with the automorphism group of the
upper half--plane, induces, when acting on the Poincar\'e
connection $A$ itself, the gauge transformation $A\to A+d\chi$,
with $\chi(z)=\ln(\mu_z/\bar\mu_z)^{i\over2}$.

\section*{Acknowledgements} It is a pleasure to thank Gaetano Bertoldi and Jos\'e Isidro for collaboration in the early stage
of this work. We also thank
 Giulio Bonelli and Dima Sorokin for interesting
discussions and Roberto Volpato for enlightening discussions and careful reading of the paper.

\end{document}